\documentclass{aa}
\usepackage{graphicx}
\usepackage{natbib}

\bibpunct{(}{)}{;}{a}{}{,}

\begin{document}

\title{Microlensing towards the Large Magellanic Cloud}
\titlerunning{Microlensing towards the LMC}
\author{Ph. Jetzer\inst{1},
        L. Mancini\inst{1,2,4},
        G. Scarpetta\inst{2,3,4}}
\offprints{Ph. Jetzer, \\
    \email{jetzer@physik.unizh.ch}}
\institute{Institut f\"{u}r Theoretische Physik der
           Universit\"{a}t Z\"{u}rich, CH-8057 Z\"{u}rich, Switzerland
           \and Dipartimento di Fisica ``E.R. Caianiello'',
           Universit\`{a} di Salerno, I-84081 Baronissi (SA), Italy
           \and International Institute for Advanced Scientific Studies,
           Vietri sul Mare (SA), Italy
           \and Istituto Nazionale di Fisica Nucleare, sez. Napoli, Italy}
\date{Received  / Accepted}

\abstract{
The nature and the location of the lenses discovered in the
microlensing surveys done so far towards the LMC remain unclear.
Motivated by these questions we compute the optical depth and
particularly the number of expected events for self-lensing for
both the MACHO and EROS2 observations. We calculate these
quantities also for other possible lens populations such as thin
and thick disk and galactic spheroid. Moreover, we estimate for
each of these components the corresponding average event duration
and mean mass using the mass moment method. By comparing the
theoretical quantities with the values of the observed events it
is possible to put some constraints on the location and the nature
of the MACHOs. Clearly, given the large uncertainties and the few
events at disposal it is not possible to draw sharp conclusions,
nevertheless we find that certainly at least 3-4 MACHO events are
due to lenses in LMC, which are most probably low mass stars, but
that hardly all events can be due to self-lensing. This
conclusions is even stronger when considering the EROS2 events,
due to their spatial distribution. The most plausible solution is
that the events observed so far are due to lenses belonging to
different intervening populations: low mass stars in the LMC, in
the thick disk, in the spheroid and possibly some true MACHOs in
the halo.

}

\maketitle

\section{Introduction}
The location and the nature of the microlensing events found so
far towards the Large Magellanic Cloud (LMC) is still a matter of
controversy. The MACHO collaboration found 13 to 17 events in 5.7
years of observations, with a mass for the lenses estimated to be
in the range $0.15 - 0.9 \, \mathrm{M_{\sun}}$  assuming a
standard spherical Galactic halo (Alcook et al. 2000a) and derived
an optical depth of $\tau= 1.2^{+0.4}_{-0.3} \times 10^{-7}$. An
analysis of the spatial distribution of the events as well as a
detailed study of the source star location done on HST images
(Alcook et al. 2001a) are both consistent with an extended lens
distribution such as the Milky Way halo, however an LMC
distribution is only slightly disfavoured. Thus this test is not
conclusive given the few events at disposal. The EROS2
collaboration (Milsztajn \& Lasserre 2001) announced the discovery
of 4 events, with an estimated average mass of 0.2
$\mathrm{M_{\sun}}$, based on three years of observation but
monitoring about twice as much stars as the MACHO collaboration.
The MACHO collaboration monitored primarily 15 deg$^2$ in the
central part of the LMC, whereas the EROS experiment covers a
larger solid angle of 64 deg$^2$ but in less crowded fields. The
EROS microlensing rate should thus be less affected by
self-lensing. This might be the reason for the fewer events seen
by EROS as compared to the MACHO experiment.

It has been argued that some if not all the events could be due to
LMC self-lensing. Indeed, several authors have already studied in
detail self-lensing by focusing, however, mainly on the value of
the microlensing optical depth. Sahu (1994) and Wu (1994)
suggested that self-lensing within the LMC could explain the
observed optical depth. This claim has been questioned by further
studies (Gould 1995; Alcock et al. 1997a). A major problem is the
uncertainty related to a precise knowledge of the shape and the
total mass of the LMC. Indeed, for instance a disaligned bar, as
suggested by Weinberg, could increase the self-lensing optical
depth (Weinberg 2000; Evans \& Kerins 2000; Zhao \& Evans 2000) as
well as assuming the LMC being much more extended along the line
of sight (Aubourg et al. 1999). These authors find then an optical
depth for self-lensing of $0.5 - 1.5 \times 10^{-7}$, thus
comparable with the measured value by the MACHO collaboration.
Other authors, instead, find lower values for the optical depth in
the range $0.5 - 8.0\times 10^{-8}$ (see for instance Gyuk et al.
2000). The discrepancy are of course due to the different adopted
models for the shape of the LMC, which are based on various
arguments and by weighing differently the observation on the LMC
star distribution which are available today.

We notice that a direct comparison of the theoretical values for
the optical depth to self-lensing and the measured one is not
straightforward. Indeed, the theoretical values are computed for a
given line of sight and then to compare with the measured value
one takes the average over the lines of sight corresponding to the
various monitored fields (Gyuk et al. 2000). On the other hand the
measured optical depth is computed assuming that it stays constant
over the monitored fields of the LMC. Such a procedure is
certainly adequate if the lenses are in the halo and thus their
optical depth does practically not vary on the size of the LMC,
however, this is no longer true when dealing with self-lensing.

Some of the events found by the MACHO team are most probably due
to self-lensing: the event MACHO-LMC-9 is a double lens with
caustic crossing (Alcock et al. 2000b) and its proper motion is
very low, thus favouring an interpretation as a double lens within
the LMC; the source star for the event MACHO-LMC-14 is double
(Alcock et al. 2001b) and this has allowed to conclude that the
lens is most probably in the LMC. The expected LMC self-lensing
optical depth due to these two events has been estimated to lie
within the range $1.1-1.8\times10^{-8}$ (Alcock et al. 2001b),
which is still below the expected optical depth for self-lensing
even when considering models giving low values for the optical
depth.

The event LMC-5 is due to a disk lens (Alcock et al. 2001c) and
indeed the lens has even been observed with the HST. The lens mass
is either $\sim 0.04 ~\mathrm{M_{\sun}}$ or in the range $0.095 -
0.13~\mathrm{M_{\sun}}$, so that it is a true brown dwarf or a
M4-5V spectral type low mass star. The other stars which have been
microlensed were also observed but no lens could be detected, thus
implying that the lens cannot be a disk star but has to be either
a true halo object or a faint star or brown dwarf in the LMC
itself.

Some work has also been done on the Small Magellanic Clouds (SMC)
where only two microlensing events have been found up to now
(Alcock et al. 1997b, 1999; Palanque-Delabrouille et al. 1998).
One was a resolved binary event, allowing the determination of the
lens distance (Alcock et al. 1999; Afonso et al. 1998; Albrow et
al. 1999), which most probably resides in the SMC itself and thus
clearly being due to self-lensing. The other event is of long
duration and a detailed analysis seems also to favour a
self-lensing interpretation (Palanque-Delabrouille et al. 1998).
It has been argued that the SMC self-lensing optical depth should
be higher than the corresponding LMC value since the SMC is
tidally disrupted, due to its interaction with the Milky Way and
the LMC, and is thus quite elongated along the line of sight to
the Milky Way (Caldwell \& Coulson 1986; Welch et al. 1987).

Thus up to now the question of the location of the observed MACHO
events is unsolved and still subject to discussion. Clearly, with
much more events at disposal one might solve this problem by
looking for instance at their spatial distribution. However, since
the MACHO collaboration data taking stopped at the end of 1999 and
the EROS experiment is still underway, but will hardly lead to a
large amount of events in the next few years, there is not much
hope to have substantial more data at disposal within the next few
years and it is thus of importance to explore further ways which
can give insight into the MACHO location using only the already
existing data. This is the main aim of this paper.

As a first point we calculate the optical depth, the average
duration and particularly the number of expected events for
self-lensing. For the latter quantity we take for the number of
monitored stars and for the exposure time the values corresponding
to the MACHO and EROS experiments, respectively. Moreover, we
compute the same quantities for MACHOs located in a thin or thick
disk or in a spheroid around our Galaxy. As a main point we
compute following the mass moment method (De Rujula et al. 1991,
1992) the average mass of the lenses for the observed events under
the assumption that the lenses are located in the LMC, the halo or
in one of the Galactic components. This allows to see whether the
assumption of having the lenses in the LMC itself or in one of the
Galactic components leads to consistent values for the masses. Our
analysis indicates that most probably not all lenses originate
from the same population but are due to different ones. Therefore,
some of the conclusions reported so far in the literature,
especially on the average mass and thus on the nature of the
lenses, have to be taken with caution. Finally, we give an estimate
of the fraction of the local dark mass density detected in form of
MACHOs in the Galactic halo.

The paper is organized as follows: in Sect. 2 we give the density
profiles of the stellar distributions for the different components
of the LMC, which will then be used in the following. In Sect. 3
we present the different equations to compute the optical depth,
the microlensing rate and the mass moments we shall use. In Sect.
4  we report the results for the LMC self-lensing, whereas in
Sect. 5 we present the values we find for the various Galactic
lens populations: thin and thick disks, spheroid and halo. The
mass of the lenses is discussed in Sect. 6. In Sect. 7, by using
different models, we determine the fraction of the local dark mass
density detected in form of MACHOs in the Galactic halo. In Sect.
8 we analyse the asymmetry in the spatial distribution of the
observed events. We conclude in Sect. 9 with a summary of our
results.

%
\section{LMC morphology, mass and density profiles}
\label{morphology}
In the last years the puzzle of the real nature of the LMC is
showing up progressively with respect to the first picture of de
Vaucouleurs (1957). The unexpected scale length substantially
larger than most irregular galaxies led to revise the LMC shape
(Bothun \& Thompson 1988). Kim et al. (1998), emphasizing the
spiral structure of the HI gas of LMC, confirmed the opinion that
LMC has to be considered, from a morphological and dynamical point
of view, more similar to a dwarf spiral than an irregular galaxy.
However, LMC does not present a well-defined center, and there are
still uncertainties and discordant measures on many important
parameters. For this reason we will use different models and vary
some parameters in specific ranges such as to get reasonable
ranges for our theoretical results.

The LMC conventional features are a disk and an off-centered bar
lying in the same plane. The significant offset of the bar in
respect of the disk (about 1 kpc) suggests a LMC dominated by dark
matter (Cioni et al. 2000), in agreement with the hypothesis of
Still (1999) that LMC is a {\it fast rotator} dwarf galaxy. There
are many lines of evidence supporting the planar geometry of LMC
(van der Marel \& Cioni 2001). Nevertheless, several authors
suggested that some LMC populations can not reside in the disk
plane (Luks \& Rohlfs 1992; Zaritsky \& Lin 1997; Zaritsky et al.
1999; Weinberg \& Nikolaev 2001) while Zhao and Evans (2000)
proposed a misaligned offset bar model.

The surface-brightness profile of the LMC is strongly exponential
along the bar and a truncation becomes apparent at a scale length
of 1.65 kpc (Bothun \& Thompson 1988). The disk seems circular,
thin, flat and seen nearly face-on with the east side closer to us
than the west side. Its plane is tilted to the plane of the sky by
an angle of roughly $30^{\circ}-45^{\circ}$ with a position angle
of $\phi_{\mathrm{d}} \simeq170^{\circ}$ (Westerlund 1997).
Following Gyuk et al. (2000), we placed the disk center at the
kinematic center of the HI gas J2000
$(l_{\mathrm{d,0}},b_{\mathrm{d,0}})=(279.7^{\circ},-33.5^{\circ})$
(Kim et al. 1998).

Thanks to the recent DENIS and 2MASS surveys, there are strong
indications that the shape of the LMC disk is not circular, but
elliptical with a small vertical scale height ($<0.5
\,\mathrm{kpc}$) (van der Marel \& Cioni 2001). The study of
near-IR surveys yields an intrinsic ellipticity of $0.31$ in the
outer parts of the LMC (van der Marel 2001). This elongation,
considerably larger than typical for disk galaxies, is in the
direction of the Galactic center and perpendicular to the
Magellanic stream, and is likely caused by the tidal force of our
Galaxy. Very recently, van der Marel et al. (2002) pointed out
that the distribution of neutral gas is not a good tracer, and
thus leads to an incorrect LMC model. Instead, using the carbon
star data, they provided an accurate measurement of the dynamical
center of the stars, which turns out to be consistent with the
center of the bar and with the position of the center of the outer
isophotes of the LMC. Thus, the disk has a center that coincides
with the center of the bar, has an elliptical shape and is thick
and flared (van der Marel et al. 2002). We will discuss a new LMC
model that takes completely into account this revised picture of
the structure and dynamics of the LMC in a forthcoming paper.
However, we do not expect the results of this paper to be
substantially modified and thus also our main conclusions, since
the reported total mass of the LMC is well in accordance with the
values we adopted.

The bar has a size of roughly $3^{\circ} \times 1^{\circ}$, a
position angle of $\phi_{\mathrm{b}} \simeq 120^{\circ}$ and its
center is located at J2000 $(l_{\mathrm{b,0}},b_{\mathrm{b,0}})=
(280.5^{\circ},-32.8^{\circ})$ (NED\footnote{Nasa/ipac
Extragalactic Database} data). The bar has boxy contours and sharp
edges, even if new morphological studies, based on
colour-magnitude diagrams extracted from DENIS catalogue, show a
bar extending over about $4^{\circ}$ as well as an elongation at
its borders that suggests the presence of spiral arms (Cioni et
al. 2000).

There is a great debate also about the LMC mass. This topic is
discussed in detail in Gyuk et al. (2000). Here, according to
several authors (e.g. Sahu 1994; Gyuk et al. 2000), we adopt
$M_{\mathrm{bar}}+M_{\mathrm{disk}}$ in the range $3-6 \times
10^{9} \mathrm{M_{\sun}}$ with $1 \leq
M_{\mathrm{disk}}/M_{\mathrm{bar}} \leq 5$.

Another important parameter is the velocity dispersion: for the
LMC stars, we adopted a velocity dispersion in the range of $15-35
\,\mathrm{km/sec}$, as it seems that there are no LMC stellar
populations with a velocity dispersion greater than $33
\,\mathrm{km/sec}$ (Hughes et al. 1991).

In the self-lensing framework, we must consider different
source/lens geometries, e.g. sources and lenses in the disk or in
the bar, otherwise sources in the disk, lenses in the bar and
viceversa. Due to the considerable uncertainties of the various
LMC components, we will consider two different models for the
description of the LMC disk and bar populations, model 1 and model
2. Each of the models are then further subdivided into two classes
$a$ and $b$ according to the set of the adopted parameters.
\begin{itemize}
\item[$\bullet$] Model $1$.
Following Gyuk et al. (2000), we consider
$M_{\mathrm{bar}}=1/4\,M_{\mathrm{disk}}$, and adopt a circular
LMC star density profile to describe the spatial distribution for
the disk population, which will be used for both the lenses and
the sources. We can write their density as a double exponential
profile
\begin{equation}\label{disk density}
\rho_{\mathrm{d}}=\frac{M_{\mathrm{d}}}{4\pi z_{\mathrm{d}}R_{\mathrm{d}}^{2}}\,
e^{-\frac{R}{R_{\mathrm{d}}}-\left|\frac{z}{z_{\mathrm{d}}}\right|},
\end{equation}
where $R_{\mathrm{d}}=1.6 \, \mathrm{kpc}$ is the radial scale
length, $z_{\mathrm{d}}=0.3 \, \mathrm{kpc}$ the vertical scale
height and $M_{\mathrm{d}}= 2.4-4.8 \times10^{9}
\,\mathrm{M_{\sun}}$ the disk mass. $R$ and $z$ are the
cylindrical coordinates of the lens/source position. For the disk
inclination $i$ we take a value in the range $30^{\circ}$ to
$40^{\circ}$.

For the bar lens and source populations we use a triaxial gaussian
density profile to describe their distribution as follows
\begin{equation}\label{bar density}
\rho_{\mathrm{b}}=\frac{M_{\mathrm{b}}}
{\left(2\pi \right)^{3/2} x_{\mathrm{b}}y_{\mathrm{b}}
z_{\mathrm{b}}} \,
e^{-\frac{1}{2} \left[\left(\frac{x}{x_{\mathrm{b}}}\right)^{2}+
\left(\frac{y}{y_{\mathrm{b}}}\right)^{2}+
\left(\frac{z}{z_{\mathrm{b}}}\right)^{2}\right]},
\end{equation}
where $M_{\mathrm{b}}= 0.6-1.2 \times 10^{9}\,
\mathrm{{M_{\sun}}}$ is the bar mass. $x$, $y$ and $z$ are the
coordinates along the principal axes of the bar. Due to the
uncertainties we used two different sets for the scale lengths
along the axes:
\begin{itemize}
\item Model $1a$ with $x_{\mathrm{b}}= 1.0 \, \mathrm{kpc}$ and
$y_{\mathrm{b}}=z_{\mathrm{b}}= 0.3 \, \mathrm{kpc}$;
\item Model $1b$ with $x_{\mathrm{b}}=1.51 \, \mathrm{kpc}$ and
$y_{\mathrm{b}}=z_{\mathrm{b}}= 0.44 \, \mathrm{kpc}$.
\end{itemize}
All other parameters in models $1a$ and $1b$ are equal and have
values as mentioned above.
\vspace{0.5 cm}
\item[$\bullet$] Model 2.
Following the results on LMC morphology by van der Marel (2001),
we consider an elliptical disk with an inclination of $35^{\circ}$
and a line-of-nodes position angle equal to $122.5^{\circ}$. Zhao
and Evans (2000) used a LMC lensable mass in the range $2.5-5.5
\times 10^{9} \mathrm{M_{\sun}}$, with the bar making up between
$25\%-50\%$ of it, suggesting thus that bar and disk could even
have the same mass. This is supported by the fact that half of the
LMC total number of stars is in the bar and this factor increases
for younger objects (Cioni et al. 2000). We take a bar size of
$4^{\circ}\times 1^{\circ}$ (Cioni et al. 2000).

Taking into account all these facts, we model the disk as
\begin{equation}
\rho_{\mathrm{d}}=\frac{M_{\mathrm{d}}}{4 \pi
x_{\mathrm{d}}y_{\mathrm{d}}z_{\mathrm{d}}}\, e^{-\left(\sqrt{
\left(\frac{x}{x_{\mathrm{d}}}\right)^2+\left(
\frac{y}{y_{\mathrm{d}}}\right)^2}\right)
-|\frac{z}{z_{\mathrm{d}}}|},
\end{equation}
where $x$, $y$ and $z$ are the coordinates along the principal
axes of the disk, and the scale lengths are
$x_{\mathrm{d}}=1.6\,\mathrm{kpc}$,
$y_{\mathrm{d}}=0.7\,x_{\mathrm{d}}=1.12\,\mathrm{kpc}$ and
$z_{\mathrm{d}}=0.3\,\mathrm{kpc}$.

For the bar density we use a quartic exponential profile, to take
into account its boxy shape:
\begin{equation}
\rho_{\mathrm{b}}=\frac{M_{\mathrm{b}}}{\pi^{3/2} \, \Gamma
\left(\frac{5}{4}\right)\, R_{\mathrm{b}}^{2}x_{\mathrm{b}}}\, e^{
-\left(\frac{R}{R_{\mathrm{b}}}\right)^{4}
-\left(\frac{x}{x_{\mathrm{b}}}\right)^{4}},
\end{equation}
where the scale lengths are $x_{\mathrm{b}}=2.1\,\mathrm{kpc}$,
$R_{\mathrm{b}}=0.58\,\mathrm{kpc}$, $\Gamma
\left(\frac{5}{4}\right)$ is the Euler gamma function and $R$ is
the radial coordinate in the orthogonal plane with respect to the
bar principal axis $x$. In order to take into account the
uncertainties in the LMC bar and disk masses, we again considered
two cases for model 2:
\begin{itemize}
\item Model $2a$ with $M_{\mathrm{bar}}=1/4\,M_{\mathrm{disk}}$ and
$M_{\mathrm{bar}}+M_{\mathrm{disk}}=3-6 \times 10^{9}\,
\mathrm{M}_{\sun}$;
\item Model $2b$ with
$M_{\mathrm{bar}}=M_{\mathrm{disk}}$ and
$M_{\mathrm{bar}}+M_{\mathrm{disk}}=3-5 \times 10^{9}\,
\mathrm{M}_{\sun}$.
\end{itemize}
All other parameters of both model $2a$ and $2b$ are equal and
have values as mentioned above.
\end{itemize}
With the help of the coordinate transformation given in Appendix A
(see also Weinberg \& Nikolaev 2000), we can easily rewrite $R$,
$x$, $y$ and $z$ as a function of the variables observer-lens
distance, $D_{\mathrm{ol}}$, observer-source distance,
$D_{\mathrm{os}}$, and of the parameters $D_{0}$, $l$, $b$,
$l_{0}$, $b_{0}$, $\phi_{\mathrm{d}}$, $\phi_{\mathrm{b}}$, $i$.
Here $D_{0}=50\,\mathrm{kpc}$ (van der Marel et al. 2002) is the
distance between the observer and the LMC, while $l$ and $b$ are
the usual galactic coordinates.

Again, following Gyuk et al. (2000) we will also consider a LMC
halo contribution, for which we adopt a spherical model with
density profile
\begin{equation}
\rho_\mathrm{h}=\rho_0 \left(1+\frac{R^{2}}{a^{2}}\right)^{-1}~,
\end{equation}
where $a\approx2\,\mathrm{kpc}$ is the LMC halo core radius (Gyuk
et al. 2000) and $\rho_0$ the central density (see Table
\ref{G-Tau}). We use an extreme model with a LMC halo mass of
$6\times 10^{9}\,\mathrm{M_{\sun}}$ within $8\,\mathrm{kpc}$ and a
velocity dispersion
$\sigma_{\mathrm{h}}=70/\sqrt{2}~\,\mathrm{km/sec}$, with a tidal
truncation radius of $11\,\mathrm{kpc}$ (Gyuk et al. 2000).
%

\section{LMC self-lensing microlensing quantities}

\subsection{Optical depth}
An important measurable quantity in a microlensing experiment is
the optical depth $\tau$, which is defined to be the probability
that at any time a random star is magnified by a lens by more than
a factor of 1.34. For MACHOs in the Galactic halo we can in good
approximation assume that all sources in the LMC are at the same
distance $D_{\mathrm{os}}=D_0$. The optical depth to gravitational
microlensing is then defined as follows
\begin{equation}\label{optical_depth_a}
\tau=\frac{4 \pi G}{c^{2}}\int _{0}^{D_{\mathrm{os}}}\rho_{\mathrm{l}} \,
\frac{D_{\mathrm{ol}}(D_{\mathrm{os}}-D_{\mathrm{ol}})}
{D_{\mathrm{os}}} \, d D_{\mathrm{ol}},
\end{equation}
where $\rho_{\mathrm{l}}$ denotes the mass density of the lenses,
for instance MACHOs in the halo.

The assumption that the sources lie all at the same distance is no
longer acceptable if we consider self-lensing. In this case we
have to integrate not only on the distance of the lenses but also
on the distance of the sources. Moreover, the LMC extension can
not be neglected, since the number density of the sources and
lenses can change substantially with the distance (Kiraga \&
Paczy\'{n}ski 1994). Assuming that the number of detectable stars
varies with the distance as $D_{\mathrm{os}}^{2\beta}$, where
$\beta$ is a parameter that takes into account the magnitude
limitations of the observation (a reasonable range for it is $-3
\leq \beta \leq -1$), the Eq. (\ref{optical_depth_a}) becomes
\begin{eqnarray}\label{optical_depth_b}
\tau&=&\frac{4 \pi G}{c^{2}}
\left(\int _{0}^{\infty}\rho_{\mathrm{s}} \,
D_{\mathrm{os}}^{2+2\beta}dD_{\mathrm{os}}\right)^{-1}\times \\& &
\int _{0}^{\infty}
\left[\int _{0}^{D_{\mathrm{os}}}\rho_{\mathrm{l}} \,
\frac{D_{\mathrm{ol}}(D_{\mathrm{os}}-D_{\mathrm{ol}})}{D_{\mathrm{os}}} \,
dD_{\mathrm{ol}}\right]
\rho_{\mathrm{s}} \, D_{\mathrm{os}}^{2+2\beta}dD_{\mathrm{os}},\nonumber
\end{eqnarray}
where $\rho_{\mathrm{s}}$ denotes the mass density of the sources.

Clearly for self-lensing the optical depth can vary substantially
with the position of the considered target field. Therefore, one
has to compute the optical depth per unit surface and then perform
an integration over the area which has been observed in order to
compare the theoretical value of the optical depth for
self-lensing with the measured one reported in the literature.

\subsection{Microlensing event rate}
To estimate the number of microlensing events with a magnification
above a certain threshold $A_{\mathrm{th}}$ (where
$A_{\mathrm{th}}=(u^{2}_{\mathrm{th}}+2)/(u_{\mathrm{th}}
\sqrt{u^{2}_{\mathrm{th}}+4})$, $A_{\mathrm{th}}=1.34$ for
$u_{\mathrm{th}}=1$), we introduce the differential number of
microlensing events
\begin{equation}
dN_{\mathrm{ev}}=N_{*}t_{\mathrm{obs}}d\Gamma,
\end{equation}
where $N_{*}$ is the total number of monitored stars during the
observation time $t_{\mathrm{obs}}$. $d\Gamma$ is the differential
rate at which a single star is microlensed by a compact object and
it is just equal to
\begin{equation}
d\Gamma=\frac{n(\vec{x})f\left(\vec{v}_{\mathrm{l}}\right)
\, d^{3}x \, d^{3}v}{dt},
\end{equation}
where the numerator on the right hand side is the number of MACHOs
with velocity in $d^{3}v=dv_{x}dv_{y}dv_{z}$ around
$\vec{v_{\mathrm{l}}}$, and located in a volume element
$d^{3}x=dxdydz$ around the position $\vec{x}$ in the microlensing
tube. $n(\vec{x})$ is the lens number density and
$f(\vec{v}_{\mathrm{l}})$ is the velocity profile (VP) of the
lenses, which is often assumed as having Gaussian shape
characterized by a mean velocity and a velocity dispersion.
Indeed, this is strictly true only for a singular isothermal
spherical model. However, the computation of the VPs (in
particular along the line-of-sight) for non spherical shapes gets
quite complicated as one has to solve the collisionless
Boltzmann-Vlasov equation. Only for few specific models, so called
power-law models (Evans \& de Zeeuw 1994), an analytical solution
has been found. In order to overcome this difficulty several
authors (van der Marel \& Franx 1993; Gerhard 1993) proposed to
write the VPs along the line-of-sight as a new set of
Gauss-Hermite moments, which arise from the expansion in terms of
corresponding orthogonal functions: the Gauss-Hermite series
\begin{equation}
f(v_{\mathrm{l}})=\frac{1}{\sqrt{2\pi\sigma^{2}}}\,
e^{-\frac{1}{2}\left(\frac{v_{\mathrm{l}}}{\sigma}\right)^{2}}
\left\{1+\sum_{j=3}^{N} h_{j} H_{j}(v_l) \right\}~.
\end{equation}
The zeroth-order term is a Gaussian and the higher order terms
measure deviations from a Gaussian. The deviations are quantified
by the Gauss-Hermite moments $h_{3},...,h_{N}$: the even
coefficients $h_{2l}$ quantify symmetric deviations, while the odd
coefficients $h_{2l+1}$ anti-symmetric deviations. The $H_j$ are
the Hermite polynomials. Detailed studies show that the difference
with respect to assuming a pure Gaussian shape, both for
elliptical and spiral galaxies, is roughly $10-20\%$ (van der
Marel et al. 1994; Dehnen \& Gerhard 1994). Moreover, the use of
the Gauss-Hermite series introduces more extra parameters. On the
other hand the uncertainties of the LMC parameters, as its total
mass, are such that the theoretical estimates for microlensing
quantities vary already widely for the allowed range of values.
One has also to consider that the velocity dispersion of the
lenses differs according to the considered stellar population.
This last fact is in any case difficult to take into account when
solving for a self-consistent solution in the framework of the
collisionless Boltzmann-Vlasov equation. Thus for these reasons we
prefer to restrict to the lowest order of the VPs and, instead,
vary the parameters, especially the velocity dispersion, in an
enough large range. Therefore, following also Han \& Gould (1995),
for the VP we take a Maxwellian distribution with dispersion
velocity $\sigma_{\mathrm{l}}$, that is
\begin{equation}\label{velocity distribution}
f(\vec{v}_{\mathrm{l}}) \, d^{3}v=\frac{1}{\pi^{\frac{3}{2}}
\sigma_{\mathrm{l}}^{3}}e^{-\frac{\left(\vec{v}_{\mathrm{l}}\right)^{2}}
{\sigma_{\mathrm{l}}^{2}}}d^{3}v.
\end{equation}
It is convenient to write
$\vec{v}_{\mathrm{l}}=\vec{v}_{\mathrm{l\perp}}+\vec{v}_{\mathrm{l\parallel}}$
and $d^{3}v=d^{2}v_{\perp}dv_{\parallel}$, since the component of
the MACHO velocity parallel to the l.o.s.,
$v_{\parallel}=v_{D_{\mathrm{ol}}}$, does not enter in the
description of the microlensing phenomenon. We will thus perform
an integration of Eq. (\ref{velocity distribution}) on this
velocity component, such as to obtain the transversal velocity
distribution (Jetzer 1994):
\begin{equation}\label{bidimensional velocity distribution}
f(\vec{v}_{\mathrm{l\perp}})=\int_{-\infty}^{\infty}f(\vec{v}_{\mathrm{l}})
\, dv_{\mathrm{l\parallel}}=
\frac{1}{\pi\sigma_{\mathrm{l}}^{2}} \,
e^{-\frac{\left(\vec{v}_{\mathrm{l\perp}}\right)^{2}}{\sigma_{\mathrm{l}}^{2}}}.
\end{equation}
Moreover, when writing the volume and the velocity elements, we
must also take into account the transverse velocities of the
source star, $\vec{v}_{\mathrm{s\perp}}$, and of the observer,
$\vec{v}_{\sun\perp}$ (here we consider the observer co-moving
with the Sun), that is the microlensing tube moves with a
transverse velocity equal to
\begin{equation}\label{tube-velocity}
\vec{v}_{\mathrm{t\perp}}=
\left(1-\frac{D_{\mathrm{ol}}}{D_{\mathrm{os}}}\right)\vec{v}_{\sun\perp}+
\frac{D_{\mathrm{ol}}}{D_{\mathrm{os}}}
\vec{v}_{\mathrm{s\perp}}.
\end{equation}
So, we can write the volume element, $d^{3}x$, as
\begin{eqnarray}
d^{3}x&=&(\vec{v}_{\mathrm{r\perp}}\cdot\hat{n})dtdS=|\vec{v}_{\mathrm{r\perp}}|
\cos{\theta}dtdldD_{\mathrm{ol}}\\
&=& v_{\mathrm{r\perp}}\cos{\theta}dtR_{\mathrm{E}}du_{\mathrm{th}}
d{\alpha}dD_{\mathrm{ol}}\nonumber,
\end{eqnarray}
where $\vec{v}_{\mathrm{r\perp}}=\vec{v}_{\mathrm{l\perp}}-
\vec{v}_\mathrm{{t\perp}}$ is the lens transverse velocity in the
rest frame of the Galaxy,
\begin{equation}\label{EinsteinRadius}
R_{\mathrm{E}}=\sqrt{\frac{4GM_{\sun}}{c^{2}}}\left(\mu\frac{D_{\mathrm{ol}}
(D_{\mathrm{os}}-D_{\mathrm{ol}})}{D_{\mathrm{os}}}\right)^{\frac{1}{2}}
\end{equation}
is the Einstein radius, while $\theta$ is the angle  between
$\vec{v}_{\mathrm{r\perp}}$ and the normal, $\hat{n}$, to the
lateral superficial element, $dS=dldD_{\mathrm{ol}}$, of the
microlensing tube, with
$dl=R_{\mathrm{E}}du_{\mathrm{th}}d{\alpha}$ being the cylindrical
segment of the tube. Here and in the following, we use solar mass
units defined as $\mu=M/M_{\sun}$, where $M$ is the lens mass. The
velocity element, $d^{2}v_{\perp}$, is given in cylindrical
coordinates by
\begin{equation}
d^{2}v_{\perp}=v_{\mathrm{r\perp}}dv_{\mathrm{r\perp}}d{\theta}.
\end{equation}
The number $\tilde N$ of MACHOs inside a surface element of the
microlensing tube becomes
\begin{equation}
d\tilde N=n(\vec{x}) f\left(\vec{v}_{\mathrm{l\perp}}\right)
v_{\mathrm{r\perp}}^{2}\cos{\theta} \,
dt \, R_{\mathrm{E}} \, du_{\mathrm{th}} \,
d{\alpha} \, dD_{\mathrm{ol}} \, dv_{\mathrm{r\perp}}d{\theta}.
\end{equation}
Therefore, the microlensing differential event rate is just (De
R\'{u}jula et al. 1991, Griest 1991)
\begin{equation}
d\Gamma=\frac{dn}{d\mu}d\mu f(\vec{v}_{\mathrm{l\perp}})
v_{\mathrm{r\perp}}^{2}\cos{\theta} \, R_{\mathrm{E}} \, du_{\mathrm{th}} \,
d{\alpha} \, dD_{\mathrm{ol}} \, dv_{\mathrm{r\perp}}d{\theta}.
\end{equation}
%
\subsection{Mass distribution}
The determination of the mass distribution of the objects which
act as lenses is one of the main aims of microlensing experiments.
We assume, as usually, that the mass distribution of the lenses is
independent of their
position in the LMC or in the Galaxy ({\it factorization
hypothesis}). So, the number density can be written as
\begin{equation}\label{number-density}
\left(\frac{dn}{d\mu}\right)d\mu =\left(\frac{dn_{0}}{d\mu
}\right) \frac{\rho_{\mathrm{lens}}}{\rho _\mathrm{{0,lens}}}
d\mu=\left(\frac{dn_{0}}{d\mu}\right) H_{\mathrm{l}}(D_{\mathrm{ol}})d\mu,
\end{equation}
where $H_{\mathrm{l}}(D_{\mathrm{ol}})=
\rho_{\mathrm{lens}}/\rho_{\mathrm{0,lens}}$,
$\rho_{\mathrm{lens}}$ and $\rho_{\mathrm{0,lens}}$ are the
density and the local density of the lens population,
respectively. The lens number density per unit mass,
$dn_{0}/d\mu$, is normalized as follows:
\begin{equation}\label{normalization}
\int_{\mu_{\mathrm{min}}}^{\mu_{\mathrm{max}}}\frac{dn_{0}}{d\mu}\mu \, d\mu
=\frac{\rho_{\mathrm{0,lens}}}{M_{\sun}},
\end{equation}
where we will assume in the following for the lenses a minimal
mass $\mu_{\mathrm{min}}=0.1$ or 0.01 and a maximal mass
$\mu_{\mathrm{max}}=10$. If all lenses have the same mass
$\mu_{0}$, we can describe the mass distribution by a delta
function
\begin{equation}\label{delta}
\frac{dn_{0}}{d\mu}=\frac{\rho_{\mathrm{0,lens}}}
{M_{\sun}\mu_{0}}\delta(\mu-\mu_{0}).
\end{equation}
Otherwise, we can use a Salpeter-type initial mass function (IMF)
\begin{equation}\label{powerlaw}
\frac{dn_{0}}{d\mu}=C\, \mu^{-\alpha},
\end{equation}
where the classical value for $\alpha$ is $2.35$ (Salpeter 1955).
Even if this value overestimates the number of low-mass stars with
$M \leq 0.5\,\mathrm{M_{\sun}}$, in general it describes very well
stars with mass above $\sim 0.3-0.5 \, \mathrm{M_{\sun}}$ in
different regions of the Milky Way and of the nearby Magellanic
Clouds. Another choice, based upon counts of M dwarfs in the
Galactic disk, is $\alpha$ equal to $0.56$ for $\mu$ in the range
$0.1$ to $0.59$ and to $2.2$ for $\mu>0.59$ (Gould et al. 1997).
In this case, the pre-factors $C_i$ are fixed by the normalization
condition
\begin{equation}\label{normalization2}
C_{1}\int_{0.1}^{0.59}\mu^{1-0.56}d\mu+C_{2}\int_{0.59}^{10}\mu^{1-2.2}d\mu
=\frac{\rho_{\mathrm{0,lens}}}{\mathrm{M}_{\sun}}.
\end{equation}
We notice that, for these IMF slopes, $\mu_{\mathrm{min}}$ might
even be lower than $0.1$. Several alternative functional forms
have been recently proposed for the low mass and brown dwarf stars
range with a $\mu_{min} \simeq 0.01$. Reid et al. (1999) indicated
a value of $\alpha$ from $1$ to $2$, based on ultracool dwarfs
discovered by DENIS and 2MASS surveys. By analyzing methane
dwarfs, Herbst et al. (1999) suggested that $\alpha\leq 0.8$ for
disk brown dwarfs. B\'{e}jar et al. (2001), studying the
substellar objects in the $\sigma$ Orionis young stellar cluster,
found that the mass spectrum increases towards lower masses with
an exponent $\alpha\approx0.8$ for $0.01 < \mu <0.2$. Given all
these uncertainties we will in the following perform the various
calculations using four different types of IMFs (see Sect.
\ref{number of events}), two with $\mu_{min}=0.1$ and two with
$\mu_{min}=0.01$.

\subsection{Mass and time moments}
Assuming that only the faint stars up to about
$\mu_\mathrm{{up}}=1$ can contribute to microlensing events, the
mass moment method (De R\'{u}jula et al. 1991; Grenacher et al.
1999) allows to extract information on the average mass of the
lenses, relating the m-th mass moment
\begin{equation}\label{mass moment}
<\mu^{m}>=\int_{\mu_{\mathrm{min}}}^{\mu_{\mathrm{up}}}
\varepsilon_{n}(\mu)\frac{dn_{0}}{d\mu}\mu^{m}d\mu,
\end{equation}
to the cumulative n-th time moment, constructed from the
observations as follows
\begin{equation}\label{experimental time moment}
<T^{n}>=\sum_{events}T^{n},
\end{equation}
with $m\equiv(n+1)/2$. $T$ is the duration of a generic
microlensing event, that is
\begin{equation}\label{duration}
T=\frac{R_{\mathrm{E}}}{|\vec{v}_{\mathrm{r\perp}}|}=\frac{R_{\mathrm{E}}}
{|\vec{v}_{\mathrm{l\perp}}-\vec{v}_{\mathrm{t\perp}}|},
\end{equation}
while $\varepsilon _{n}(\mu)$ is the efficiency function (see De
R\'ujula et al. 1991). In this way, with the knowledge of the
event durations, we will be able to estimate the mean lens mass.

The theoretical expression for the time moments is
\begin{equation}\label{time moment}
<T^{n}>=\int\varepsilon_{n}(\mu)T^{n}dN_\mathrm{{ev}}.
\end{equation}
Clearly under the hypothesis (\ref{number-density}) the Eq.
(\ref{time moment}) factorizes. The rate of microlensing events is
given by:
\begin{eqnarray}\label{gamma}
&\Gamma&=\frac{1}{\pi\sigma_{\mathrm{l}}^{2}}\sqrt{\frac{4GM_{\sun}}{c^{2}}}
\int_{0}^{\infty}v_{\mathrm{r\perp}}^{2}
e^{-\left(\frac{\vec{v}_{\mathrm{r\perp}}+\vec{v}_{\mathrm{t\perp}}}
{\sigma_{\mathrm{l}}}\right)^{2}}dv_{\mathrm{r\perp}} \nonumber \\& &
\times \int_{0}^{D_{\mathrm{os}}}
\sqrt{\frac{D_{\mathrm{ol}}(D_{\mathrm{os}}-D_{\mathrm{ol}})}
{D_{\mathrm{os}}}}H_{\mathrm{l}}(D_{\mathrm{ol}})dD_{\mathrm{ol}}
\int_{0}^{2 \pi}d\alpha
\\& &
\times
\int_{-\pi/2}^{\pi/2}\cos{\theta} \, d\theta
\int_{0}^{u_{\mathrm{min}}}du_{\mathrm{th}}
\int_{\mu_{\mathrm{min}}}^{\mu_{\mathrm{up}}} \varepsilon_0(\mu)
\sqrt{\mu} \, \frac{dn_{0}}{d\mu} \, d\mu. \nonumber
\end{eqnarray}
The $\theta$ integration is between $(-\pi/2,\pi/2)$, since we are
only interested in lenses entering the microlensing tube.

As we already mentioned in the above section, since we are
considering LMC self-lensing, the Eq. (\ref{gamma}) has to be
integrated not only over the distance of the lenses but also over
the distance of the sources.

Till now we have considered the most general situation. However,
because the distance among lenses and sources in the self-lensing
hypothesis is small when compared with the observer-lens or
observer-source distance, we have that
$\frac{D_{\mathrm{ol}}}{D_{\mathrm{os}}} \approx1$ (Alves \&
Nelson 2000). Therefore, the transverse tube velocity, as defined
in Eq. (\ref{tube-velocity}), is roughly equal to the source
velocity,
$\vec{v}_{\mathrm{t\perp}}\approx\vec{v}_{\mathrm{s\perp}}$, and
the integral in $d\alpha$ becomes trivial. Indeed, we have checked
numerically that the results with or without this approximation
differ very little, so that the induced error is negligible.

Moreover, we must also take into account the fact that the source
stars are not at rest but have a velocity $\vec{v}_{\mathrm{s}}$.
Since we consider that both the lenses and sources are in the LMC,
we can use another Maxwellian to describe the velocity
distribution of the sources, with the same velocity dispersion
($\sigma_{\mathrm{s}}=\sigma_{\mathrm{l}}=\sigma$) as for the
lenses:
\begin{equation}\label{vs_distribution}
\bar{F}(\vec{v}_{\mathrm{s\perp}})dv_{\mathrm{s\perp}}d\varphi=\frac{v_{\mathrm{s\perp}}}
{\pi \sigma_{\mathrm{s}}^{2}}
e^{\frac{v_{\mathrm{s\perp}}^{2}}{\sigma_{\mathrm{s}}^{2}}}dv_{\mathrm{s\perp}}d\varphi,
\end{equation}
where $\varphi$ is the angle between $\vec{v}_{\mathrm{r\perp}}$
and $\vec{v}_{\mathrm{s\perp}}$. Therefore, taking into account
all these facts, the microlensing rate becomes
\begin{eqnarray}\label{microlensing-rate}
\Gamma&=&\frac{4 \, u_{\mathrm{min}}}{\pi\sigma^{4}}
\sqrt{\frac{4GM_{\sun}}{c^{2}}}
\frac{I_{\mathrm{D}}^{\frac{1}{2}}}{H_{0}}
\int_{\mu_{\mathrm{min}}}^{\mu_{\mathrm{up}}}\varepsilon_0(\mu)
\sqrt{\mu}\, \frac{dn_{0}}{d\mu}d\mu
\int_{0}^{2\pi}d\varphi\nonumber
\\
&\times&\int_{0}^{\infty}v_{\mathrm{s\perp}}
e^{-\frac{v_{\mathrm{s\perp}}^{2}}{\sigma_{\mathrm{s}}^{2}}}dv_{\mathrm{s\perp}}
\int_{0}^{\infty}v_{\mathrm{r\perp}}^{2}
e^{-\left(\frac{\vec{v}_{\mathrm{r\perp}}+\vec{v}_{\mathrm{s\perp}}}
{\sigma_{\mathrm{l}}}\right)^{2}}
dv_{r\perp},
\end{eqnarray}
where
\begin{eqnarray}
I_{\mathrm{D}}^{\frac{1}{2}}&=& \int_{0}^{\infty}
\left[\int_{0}^{D_{\mathrm{os}}}\sqrt{\frac{D_{\mathrm{ol}}
(D_{\mathrm{os}}-D_{\mathrm{ol}})}{D_{\mathrm{os}}}}H_{\mathrm{l}}
(D_{\mathrm{ol}})dD_{\mathrm{ol}}\right] \nonumber\\&\times &
H_{\mathrm{s}}(D_{\mathrm{os}})\,
D_{\mathrm{os}}^{2+2\beta}dD_{\mathrm{os}},
\end{eqnarray}
with
$H_{0}=\int_{0}^{\infty}H_{\mathrm{s}}(D_{\mathrm{os}})D_{\mathrm{os}}^{2+2\beta}
dD_{\mathrm{os}}$ and
$H_{\mathrm{s}}(D_{\mathrm{os}})=\rho_{\mathrm{s}}
/\rho_{\mathrm{0,s}}$. Then, adopting for the constant $\beta$ the
value $\beta=-1$ (which should be reasonable when considering main
sequence stars), the general expression for the n-th time moment
is:
\begin{eqnarray}
<T^{n}>&=&\int\varepsilon_{n}(\mu)T^{n}dN_{\mathrm{ev}}=\nonumber  \\&= &
\int\varepsilon_{n}(\mu)
\left(\frac{R_{\mathrm{E}}}{v_{\mathrm{r\perp}}}\right)^{n}dN_{\mathrm{ev}}=  \\&= &
\frac{4k}{\pi H_{0}} <\mu^{m}> \,  I_{\mathrm{v}}^{n}\, I_{\mathrm{D}}^{m}
\left(\frac{4GM_{\sun}}{c^{2}}\right)^{m}\nonumber,
\end{eqnarray}
with
\begin{equation}
I_{\mathrm{v}}^{n}=
\frac{2 \pi}{\sigma^{n-1}}
\int_{0}^{\infty}\int_{0}^{\infty}
I_{0} \left(2 \eta_{1} \eta_{2} \right)
\eta_{1} \eta_{2}^{2-n}
e^{-2\eta_{1}^{2}-\eta_{2}^{2}}
d\eta_{1} d\eta_{2},
\end{equation}
where $\eta_{1}=v_{\mathrm{s\perp}}/\sigma$,
$\eta_{2}=v_{\mathrm{r\perp}}/\sigma$, $I_{0}$ is the modified
Bessel function of order $0$,
$k=u_{\mathrm{min}}N_{\star}t_{\mathrm{obs}}$, and
\begin{eqnarray}
I_{\mathrm{D}}^{m}&=&\int_{0}^{\infty}
\left[\int_{0}^{D_{\mathrm{os}}}
\left(\frac{D_{\mathrm{ol}}
(D_{\mathrm{os}}-D_{\mathrm{ol}})}{D_{\mathrm{os}}}\right)^{m}
H_{\mathrm{l}}(D_{\mathrm{ol}})dD_{\mathrm{ol}}\right]
\nonumber \\&\times &
H_{\mathrm{s}}(D_{\mathrm{os}})\, dD_{\mathrm{os}}.
\end{eqnarray}
We are able now to write the time moments
$<T^{0}>$, $<T^{1}>$ and  $<T^{-1}>$:
\begin{eqnarray}
&<T^{0}>&=\sqrt{\frac{4GM_{\sun}}{c^{2}}}<\mu^{\frac{1}{2}}>
\gamma\left(1/2\right),
\\ &<T^{1}>&=\frac{4GM_{\sun}}{c^{2}}<\mu^{1}>
\gamma\left(1\right),
\\ &<T^{-1}>&=<\mu^{0}>
\gamma\left(0\right),
\end{eqnarray}
where
\begin{equation}
\gamma\left(1/2\right)=
\frac{4k}{\pi H_{0}}\, I_{\mathrm{v}}^{0}\,
I_{\mathrm{D}}^{1/2},
\end{equation}
\begin{equation}
\gamma\left(1\right)=
\frac{4k}{\pi H_{0}}\, I_{\mathrm{v}}^{1}\,
I_{\mathrm{D}}^{1},
\end{equation}
\begin{equation}
\gamma\left(0\right)=
\frac{4k}{\pi H_{0}}\, I_{\mathrm{v}}^{-1}\,
I_{\mathrm{D}}^{0}.
\end{equation}
The microlensing quantities of interest are the average event duration
\begin{equation}\label{average-event-duration}
<T>=\frac{<T^{1}>}{<T^{0}>}=
\sqrt{\frac{4G\mathrm{M}_{\sun}}{c^{2}}} \frac
{<\mu^{1}>\gamma\left(1\right)}
{<\mu^{\frac{1}{2}}>\gamma\left(\frac{1}{2}\right)},
\end{equation}
and the lens mean mass
\begin{equation}\label{mean-mass}
<M>=\frac{<\mu^{1}>}{<\mu^{0}>}=
\left(
\frac
{4GM_{\sun}}
{c^{2}}
\right)^{-1}
\frac
{<T^{1}>\gamma(0)}
{<T^{-1}>\gamma(1)}.
\end{equation}
For the mean mass, according to Eq. (\ref{experimental time
moment}), the time moments $<T^{1}>$ and $<T^{-1}>$ are calculated
using the experimental values.

\subsection{Relation among $<T>$, $\Gamma$ and $\tau$}
The primary observables in a microlensing experiment are the
optical depth $\tau$, the rate $\Gamma$ and the average duration
of the events $<T>$. We saw that the duration of a particular
microlensing event is $T=R_{\mathrm{E}}/v_{\mathrm{r\perp}}$ and
its average, Eq. (\ref{average-event-duration}), gives the time
scale of the microlensing events
\begin{equation}
<T>=\frac{<T^{1}>}{<T^{0}>}=\frac{
\int{T \, \varepsilon_{1}(\mu)dN_{ev}}}{\int\varepsilon_{0}(\mu)dN_{ev}}
=\frac{
\int{T \, \varepsilon_{1}(\mu)d\Gamma}}{\int\varepsilon_{0}(\mu)d\Gamma}.
\end{equation}
Assuming $\varepsilon_{1}(\mu)=\varepsilon_{0}(\mu)=constant$, we
have that the time scale is then equal to
$<T>=(\int{T\,d\Gamma})/\Gamma$. Recalling
Eqs.(\ref{optical_depth_b}) and (\ref{microlensing-rate}), the
relation among the three observables is just
\begin{equation}
<T>=\frac{2\tau}{\pi \Gamma}\frac{M_{\sun}}{\rho_{\mathrm{0,lens}}}<\mu^{1}>,
\end{equation}
where we have set $u_{\mathrm{min}}=1$. If $\mu_{\mathrm{up}}$ is
equal to $\mu_{\mathrm{max}}$, the ratio
$\frac{M_{\sun}}{\rho_{\mathrm{0,lens}}}<\mu^{1}>=1$ (see Eq.
(\ref{normalization})), and we get the known relation
$<T>=\frac{2\tau}{\pi \Gamma}$.

%
\section{Estimate of microlensing quantities for LMC self-lensing}
With the formalism developed in the previous sections, we are able
to get the theoretical estimates for several important
microlensing quantities. These estimates are obtained using the
values for the parameters as described in Sect. \ref{morphology}
(only the velocity dispersion and the experimental efficiency are
left unspecified, although we will also consider the influence on
our results of these parameters). While the uncertainties of the
scale lengths and the LMC disk inclination induce only minor
modifications in our results, the larger uncertainties on the LMC
mass are much more important. All results are summarized in Tables
\ref{Tau-1} and \ref{Nev-T} for the models $1a$ and $1b$, and in
Tables \ref{Tau-2} and \ref{Nev-2} for the models $2a$ and $2b$.
The LMC mass (without the halo) is in the range $3-6 \times10^{9}
\, \mathrm{M}_{\sun}$. Four different IMFs were used as specified
below. We notice also that using the above discussed source and
lens distributions for the LMC we did not take into account that
there are also gas and dust clouds which will obscure some regions
of the LMC and thus due to that the values for the optical depth
and especially for the expected number of self-lensing events have
to be considered as upper limits.
%

\subsection{Optical depth}
\begin{figure}
 \resizebox{\hsize}{!}{\includegraphics{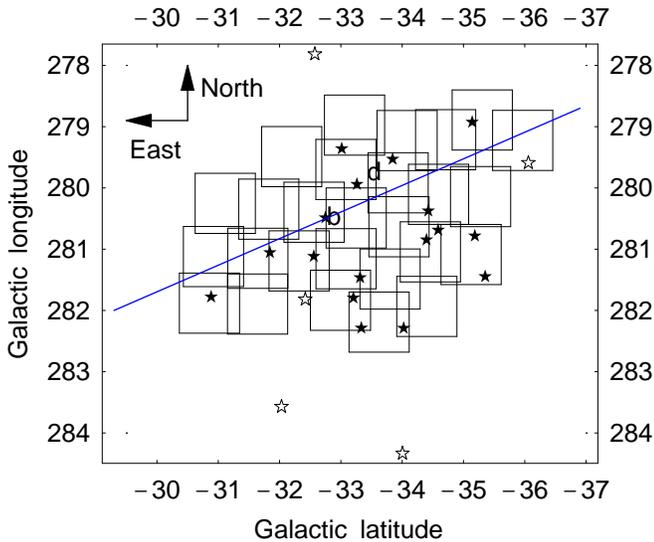}}
 \caption{The 27 MACHO fields that we selected. The filled stars
denote the positions of the microlensing events reported by the
MACHO collaboration, while the empty stars are that of the EROS
collaboration. The centers of the LMC bar and disk are labeled by
``b'' and ``d'', respectively. The blue line marks the LMC bar
major axis.}
 \label{Mfields}
\end{figure}

We computed using Eq. (\ref{optical_depth_b}) the optical depth
for several different source/lens geometries. For each of them,
adopting the parameters discussed in Sect. \ref{morphology}, we
calculated the optical depth in the directions of $27$ MACHO
fields, see Fig. \ref{Mfields}.
These fields (1, 2, 3, 5, 6, 7, 9, 10, 11, 12, 13, 14, 15, 17, 18,
19, 22, 23, 24, 47, 76, 77, 78, 79, 80, 81, 82 according to the
MACHO numbering) have been selected taking into account the
position of the microlensing events found towards the LMC by the
MACHO collaboration during a $5.7$ years monitoring campaign
(Alcock et al. 2000a).

Following the strategy of Gyuk et al. (2000), the values of the
optical depths, found for each source/lens geometry, have been
averaged on the 27 MACHO fields that we selected. The value for
each field was obtained by computing the optical depth per unit
surface and then by integrating over the area of the field.
\begin{itemize}
\item[$\bullet$]Model 1.
The total optical depth, obtained using the ``preferred'' values
for $M_{\mathrm{disk}}+M_{\mathrm{bar}}=3 \times
10^9\,\mathrm{M}_{\sun}$ and an inclination angle of $30^{\circ}$,
is $\tau=2.63 \times 10^{-8}$ for the model $1a$ and $\tau=2.58
\times 10^{-8}$ for the model $1b$. The results, summarized in
Table \ref{Tau-1}, are in good agreement with the corresponding
values found by Gyuk et al. (2000) as expected, since we adopted
the same values for the LMC parameterization. There is a slight
dependence on the LMC disk inclination, as it appears also from
Fig. \ref{Tau-inclination}, which shows the variation of the total
optical depth with the disk inclination.
\begin{figure}
 \resizebox{\hsize}{!}{\includegraphics{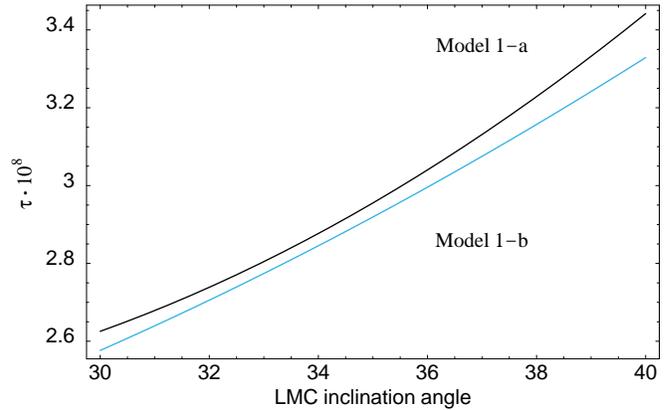}}
 \caption{Total optical depth for LMC self-lensing as a function
 of the LMC inclination angle for two different models assuming
 a total mass of $3\times 10^9\,\mathrm{M}_{\sun}$.}
 \label{Tau-inclination}
\end{figure}
If we use the upper limit for the LMC disk and bar masses, the
corresponding total optical depth is clearly larger: $\tau=5.25
\times 10^{-8}$ for the model $1a$ and $\tau=5.15 \times 10^{-8}$
for the model $1b$. By appropriately choosing the parameters in
the ranges specified in Sect. 2 Gyuk et al. (2000) found an upper
value of about $\tau=8 \times 10^{-8}$ for the self-lensing
optical depth. Clearly, this implies to adopt rather unrealistic
values for the parameters. We refer to Gyuk et al. (2000) for a
detailed discussion on the self-lensing optical depth range as
well as a comparison with the theoretical results appeared
meanwhile in the literature.
\begin{table*}
\centering \caption{Model 1: theoretical estimates of the optical
depth, $\tau$, towards the LMC, averaged over 27 MACHO fields, for
different source/lens geometries and parameters in the case of
self-lensing.}
\begin{tabular}[h]{cccccc}
\hline \hline
$Source/Lens$&$LMC \, mass$&$Model$&$LMC$&$Relative$&$\tau$\\
$Geometry$&$(\times10^{9}M_{\sun})$&$$&$inclination$&$Weight$&$(\times10^{-8})$\\
\hline $Disk/disk$&$3-6$&$\begin{tabular}{c}
1a  \\
1a  \\
1a  \\
1b  \\
1b  \\
1b  \\
\end{tabular}$&$\begin{tabular}{c}
$30^{\circ}$ \\
$35^{\circ}$ \\
$40^{\circ}$ \\
$30^{\circ}$ \\
$35^{\circ}$ \\
$40^{\circ}$ \\
\end{tabular}$&
$\begin{tabular}{c}
0.40 \\
0.35\\
0.31 \\
0.69\\
0.64 \\
0.60\\
\end{tabular}$&$\begin{tabular}{c}
1.27 - 2.54 \\
1.38 - 2.76\\
1.52 - 3.04\\
1.27 - 2.54 \\
1.38 - 2.76\\
1.52 - 3.04\\
\end{tabular}$ \\
\hline
$Disk/bar$&$3-6$&$\begin{tabular}{c}
1a  \\
1a  \\
1a  \\
1b  \\
1b  \\
1b  \\
\end{tabular}$&$\begin{tabular}{c}
$30^{\circ}$ \\
$35^{\circ}$ \\
$40^{\circ}$ \\
$30^{\circ}$ \\
$35^{\circ}$ \\
$40^{\circ}$ \\
\end{tabular}$&
$\begin{tabular}{c}
0.40 \\
0.35\\
0.31 \\
0.69\\
0.64 \\
0.60\\
\end{tabular}$&$\begin{tabular}{c}
0.83 - 1.66 \\
0.74 - 1.47\\
0.66 - 1.32\\
0.80 - 1.59 \\
0.78 - 1.56\\
0.71 - 1.42\\
\end{tabular}$ \\
\hline $Bar/disk$&$3-6$&$\begin{tabular}{c}
1a  \\
1a  \\
1a  \\
1b  \\
1b  \\
1b  \\
\end{tabular}$&$\begin{tabular}{c}
$30^{\circ}$ \\
$35^{\circ}$ \\
$40^{\circ}$ \\
$30^{\circ}$ \\
$35^{\circ}$ \\
$40^{\circ}$ \\
\end{tabular}$&
$\begin{tabular}{c}
0.60 \\
0.65\\
0.69 \\
0.31\\
0.36 \\
0.40\\
\end{tabular}$&$\begin{tabular}{c}
1.20 - 2.40 \\
1.50 - 2.94\\
1.79 - 3.59\\
2.31 - 4.62 \\
2.76 - 5.53\\
3.29 - 6.57\\
\end{tabular}$ \\
\hline $Bar/bar$&$3-6$&$\begin{tabular}{c}
1a  \\
1a  \\
1a  \\
1b  \\
1b  \\
1b  \\
\end{tabular}$&$\begin{tabular}{c}
$30^{\circ}$ \\
$35^{\circ}$ \\
$40^{\circ}$ \\
$30^{\circ}$ \\
$35^{\circ}$ \\
$40^{\circ}$ \\
\end{tabular}$&
$\begin{tabular}{c}
0.60 \\
0.65\\
0.69 \\
0.31\\
0.36 \\
0.40\\
\end{tabular}$&$\begin{tabular}{c}
1.78 - 3.58 \\
1.91 - 3.84\\
2.21 - 4.43\\
1.38 - 2.76 \\
1.50 - 3.00\\
1.65 - 3.30\\
\end{tabular}$ \\
\hline $Total$&$3-6$&$\begin{tabular}{c}
1a  \\
1a  \\
1a  \\
1b  \\
1b  \\
1b  \\
\end{tabular}$&$\begin{tabular}{c}
$30^{\circ}$ \\
$35^{\circ}$ \\
$40^{\circ}$ \\
$30^{\circ}$ \\
$35^{\circ}$ \\
$40^{\circ}$ \\
\end{tabular}$&
$\begin{tabular}{c}
1 \\
1 \\
1 \\
1 \\
1 \\
1 \\
\end{tabular}$&$\begin{tabular}{c}
2.63 - 5.25 \\
2.99 - 5.88 \\
3.44 - 6.88 \\
2.58 - 5.15 \\
2.92 - 5.84 \\
3.33 - 6.66 \\
\end{tabular}$ \\
\hline
\end{tabular}
\label{Tau-1}
\end{table*}

\vspace{0.5 cm}
\item[$\bullet$] Model 2.
The total optical depth is $\tau=3.46 \times 10^{-8}$ for the
model $2a$ and $\tau=4.05 \times 10^{-8}$ for the model $2b$ for
the lower mass value. These values are quite similar to the ones
found for model 1. The adopted parameters are described in Sect.
\ref{morphology} and the results are summarized in Table
\ref{Tau-2} for different source/lens geometries. As in the
previous case, if we use the upper limit for the LMC disk and bar
masses, the corresponding total optical depth gets larger and
probably unrealistic for both models. Nonetheless, even in that
case the observed value for the optical depth cannot be explained
only by self-lensing.
\end{itemize}
\begin{table*}
\centering \caption{Model 2: theoretical estimates of the optical
depth, $\tau$, towards the LMC, averaged over 27 MACHO fields, for
different source/lens geometries and parameters in the case of
self-lensing.}
\begin{tabular}[h]{ccccc}
\hline \hline
$Source/Lens$&$Model$&$LMC \,mass$&$Relative$&$\tau$\\
$Geometry$&$$&$(\times10^{9}\mathrm{M_{\sun}})$&$Weight$&$(\times10^{-8})$\\
\hline $Disk/disk$&$\begin{tabular}{c}
2a  \\
2b  \\
\end{tabular}$&$\begin{tabular}{c}
3 - 6 \\
3 - 5 \\
\end{tabular}$&$\begin{tabular}{c}
0.50 \\
0.20 \\
\end{tabular}$&$\begin{tabular}{c}
2.14 - 4.61 \\
1.34 - 2.40 \\
\end{tabular}$ \\
\hline
$Disk/bar$&$\begin{tabular}{c}
2a  \\
2b  \\
\end{tabular}$&$\begin{tabular}{c}
3 - 6 \\
3 - 5 \\
\end{tabular}$&$\begin{tabular}{c}
0.50 \\
0.20 \\
\end{tabular}$&$\begin{tabular}{c}
1.79 - 3.59 \\
4.48 - 7.47 \\
\end{tabular}$ \\
\hline
$Bar/disk$&$\begin{tabular}{c}
2a  \\
2b  \\
\end{tabular}$&$\begin{tabular}{c}
3 - 6 \\
3 - 5 \\
\end{tabular}$&$\begin{tabular}{c}
0.50 \\
0.80 \\
\end{tabular}$&$\begin{tabular}{c}
1.31 - 2.63 \\
0.82 - 1.36 \\
\end{tabular}$ \\
\hline
$Bar/bar$&$\begin{tabular}{c}
2a  \\
2b  \\
\end{tabular}$&$\begin{tabular}{c}
3 - 6 \\
3 - 5 \\
\end{tabular}$&$\begin{tabular}{c}
0.50 \\
0.80 \\
\end{tabular}$&$\begin{tabular}{c}
1.68 - 3.36 \\
2.79 - 6.99 \\
\end{tabular}$ \\
\hline
$Total$&$\begin{tabular}{c}
2a  \\
2b  \\
\end{tabular}$&$\begin{tabular}{c}
3 - 6 \\
3 - 5 \\
\end{tabular}$&$\begin{tabular}{c}
1 \\
1 \\
\end{tabular}$&$\begin{tabular}{c}
3.46 - 7.10 \\
4.05 - 8.65 \\
\end{tabular}$ \\
\hline
\end{tabular}
\label{Tau-2}
\end{table*}

The five events found so far by the EROS collaboration are all
located quite far from the LMC central regions, see Fig.
\ref{Mfields}. Therefore, we expect the optical depth for
self-lensing to be accordingly small. Indeed, using the model
$1a$, the averaged self-lensing optical depth over the event
positions is $0.8 \times 10^{-8}$ for a LMC disk mass equal to
$2.4\times 10^{9}\, \mathrm{M_{\sun}}$ or $1.6 \times 10^{-8}$ for
a LMC disk mass equal to $4.8\times 10^{9}\,\mathrm{M_{\sun}}$
(given the position of the EROS events the bar does not
contribute). Clearly, these values are smaller than the
corresponding ones for the MACHO fields and thus the EROS events
are less likely to be due to self-lensing.

In Table \ref{G-Tau} we give also the value for the optical depth
due to MACHOs located in the LMC halo assuming a core radius $a$
equal to 2 kpc (it was found that the optical depth is quite
insensitive to the variation of the core radius (Gyuk et al.
2000)) and that MACHOs make up 100\% of the dark LMC halo, which
is clearly an extreme assumption and has thus to be considered as
an upper limit given also the fact that it is still not clear if
the LMC has a halo at all. Moreover, this value is valid for a
line of sight going through the center of the LMC and, therefore,
also for that reason it has to be considered as an upper limit.
%

\subsection{Number of events}\label{number of events}
\begin{figure}
 \resizebox{\hsize}{!}{\includegraphics{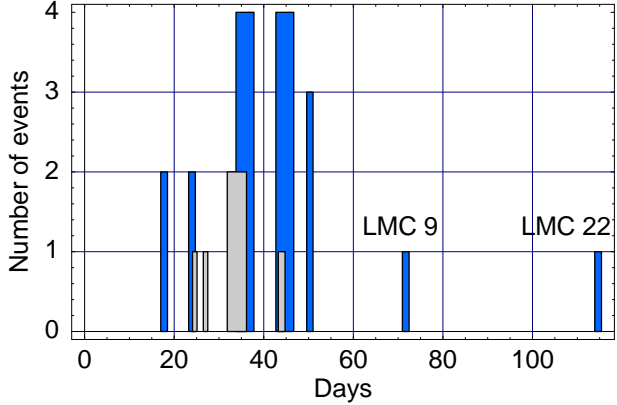}}
 \caption{Bar chart of the number of observed microlensing events
towards the LMC as a function of their duration. The full bars
represent the seventeen MACHO events, while the gray bars
represent the five EROS events.}
 \label{istogramma}
\end{figure}

By observing roughly 12 millions stars from 1992 to 1998, the
MACHO collaboration found 13-17 microlensing events, according to
the data-analysis criteria used ($A$ or $B$) (Alcock et al.
2000a). Instead, the EROS2 collaboration found 4 events monitoring
about 25.5 millions of stars from 1996 to 1999 (Milsztajn \&
Lasserre 2001). Another event was found by EROS1 in the period
1990-1993 monitoring roughly 8 milion of stars (Aubourg et al.
1993). These results are summarized in the bar chart of Fig.
\ref{istogramma}, from which one sees that the detection of events
with short duration ($\hat{t}<60 \, \mathrm{days}$) is favoured.
Yet, the two MACHO events, LMC-9 and LMC-22, both with $\hat{t}>60
\, \mathrm{days}$,  must be treated with caution, because they are
a binary event (Alcock et al. 2000b) and a possible exotic event
(Alcock et al. 2000a), respectively.

We computed the number of self-lensing events, $N_{\mathrm{ev}}$,
again by averaging over the 27 MACHO field as done for the optical
depth. We used also the MACHO parameters:
$N_{*}=11.9\times10^{6}\, \mathrm{stars}$,
$t_{obs}=5.7\,\mathrm{years}$, $\varepsilon=0.4-0.5$. We recall
that the number of events is
$N_{\mathrm{ev}}=N_{*}t_{\mathrm{obs}}\Gamma$, where $\Gamma$ is
given by Eq. (\ref{microlensing-rate}). For the mass distribution
we used four different Salpeter-type IMFs as given in Eq.
(\ref{powerlaw}), with, as an illustration, different values for
$\alpha$ and $\mu_{min}$:
\begin{itemize}
\item[$i)$] IMF-1 with $\mu_{\mathrm{min}}=0.1$ and
$\alpha=0.56$ for $0.1<\mu<0.59$ and
$\alpha=2.2$ for $\mu>0.59$;
\item[$ii)$] IMF-2 with $\mu_{\mathrm{min}}=0.1$ and $\alpha=2.35$;
\item[$iii)$] IMF-3 with $\mu_{\mathrm{min}}=0.01$ and $\alpha=0.8$
for $0.01<\mu<0.2$ and $\alpha=2.35$ for $\mu>0.2$;
\item[$iv)$] IMF-4 with $\mu_{\mathrm{min}}=0.01$ and $\alpha=2.35$.
\end{itemize}
\begin{itemize}
\item[$\bullet$]Model 1.
The results for the different source/lens geometries, IMFs and
parameters are shown in Table \ref{Nev-T} for $\varepsilon=0.5$,
$\sigma=30\,\mathrm{km/sec}$. For other values of $\varepsilon$
and $\sigma$ it is sufficient to rescale the numbers in the table
by the new values of the parameters, since the number of events is
proportional to $\varepsilon$ and $\sigma$.
\begin{table*}
\centering \caption{Model 1: theoretical estimates of the LMC
self-lensing number of events, $N_{\mathrm{ev}}$, averaged over 27
MACHO fields. We used $\varepsilon=0.5$ for the detection
efficiency and $\sigma=30\,\mathrm{km \, sec^{-1}}$ for the
velocity dispersion. Different source/lens geometries and four
Salpeter-type IMFs with different slopes and minimal lens mass as
described in the text were used.}
\begin{tabular}[h]{ccccccccc}
\hline \hline $Source/Lens$&$LMC \,
mass$&$Model$&$LMC$&$Relative$&$N_{\mathrm{ev}}$
&$N_{\mathrm{ev}}$&$N_{\mathrm{ev}}$&$N_{\mathrm{ev}}$\\
$Geometry$&$(\times10^{9}\mathrm{M_{\sun}})$&$$&$inclination$&$Weight$&
$(\mathrm{IMF-1})$&$(\mathrm{IMF-2})$&$(\mathrm{IMF-3})$&$(\mathrm{IMF-4)}$ \\
\hline $Disk/disk$&$3-6$&$\begin{tabular}{c}
1a  \\
1a  \\
1b  \\
1b  \\
\end{tabular}$&$\begin{tabular}{c}
$30^{\circ}$ \\
$35^{\circ}$ \\
$30^{\circ}$ \\
$35^{\circ}$ \\
\end{tabular}$&
$\begin{tabular}{c}
0.40 \\
0.35\\
0.69\\
0.64 \\
\end{tabular}$&$\begin{tabular}{c}
0.60-1.20 \\
0.60-1.35  \\
0.60-1.20  \\
0.60-1.35  \\
\end{tabular}$&$\begin{tabular}{c}
1.50-2.85  \\
1.50-3.00  \\
1.50-2.86  \\
1.50-3.00  \\
\end{tabular}$&$\begin{tabular}{c}
1.62-3.23  \\
1.72-3.44  \\
1.62-3.23 \\
1.72-3.44 \\
\end{tabular}$&$\begin{tabular}{c}
4.35-8.85  \\
4.65-9.45  \\
4.35-8.85  \\
4.65-9.45  \\
\end{tabular}$ \\
\hline
$Disk/bar$&$3-6$&$\begin{tabular}{c}
1a  \\
1a  \\
1b  \\
1b  \\
\end{tabular}$&$\begin{tabular}{c}
$30^{\circ}$ \\
$35^{\circ}$ \\
$30^{\circ}$ \\
$35^{\circ}$ \\
\end{tabular}$&
$\begin{tabular}{c}
0.40 \\
0.35\\
0.69\\
0.64 \\
\end{tabular}$&$\begin{tabular}{c}
0.45-1.05 \\
0.45-0.75 \\
0.15-0.30 \\
0.15-0.15 \\
\end{tabular}$&$\begin{tabular}{c}
1.20-2.25 \\
0.90-1.80 \\
0.30-0.60 \\
0.30-0.45 \\
\end{tabular}$&$\begin{tabular}{c}
1.31-2.63 \\
1.04-2.08 \\
0.33-0.66 \\
0.30-0.60 \\
\end{tabular}$&$\begin{tabular}{c}
3.60-7.20 \\
2.85-5.70 \\
0.90-1.80 \\
0.75-1.65 \\
\end{tabular}$ \\
\hline
$Bar/disk$&$3-6$&$\begin{tabular}{c}
1a  \\
1a  \\
1b  \\
1b  \\
\end{tabular}$&$\begin{tabular}{c}
$30^{\circ}$ \\
$35^{\circ}$ \\
$30^{\circ}$ \\
$35^{\circ}$ \\
\end{tabular}$&
$\begin{tabular}{c}
0.60 \\
0.65 \\
0.31 \\
0.36 \\
\end{tabular}$&$\begin{tabular}{c}
0.90-1.80 \\
1.35-2.55 \\
2.25-4.50 \\
2.85-5.70 \\
\end{tabular}$&$\begin{tabular}{c}
2.10-4.35 \\
3.00-6.00 \\
5.10-10.2 \\
6.60-13.1 \\
\end{tabular}$&$\begin{tabular}{c}
2.50-5.00 \\
3.47-6.94 \\
5.95-11.9 \\
7.57-15.1 \\
\end{tabular}$&$\begin{tabular}{c}
6.75-13.7 \\
9.45-19.1 \\
16.4-32.6 \\
20.7-41.4\\
\end{tabular}$ \\
\hline
$Bar/bar$&$3-6$&$\begin{tabular}{c}
1a  \\
1a  \\
1b  \\
1b  \\
\end{tabular}$&$\begin{tabular}{c}
$30^{\circ}$ \\
$35^{\circ}$ \\
$30^{\circ}$ \\
$35^{\circ}$ \\
\end{tabular}$&
$\begin{tabular}{c}
0.60 \\
0.65\\
0.31\\
0.36 \\
\end{tabular}$&$\begin{tabular}{c}
1.05-2.10 \\
1.20-2.25 \\
0.75-1.35 \\
0.75-1.50 \\
\end{tabular}$&$\begin{tabular}{c}
2.40-4.80 \\
2.55-5.35 \\
1.50-3.15 \\
1.65-3.45 \\
\end{tabular}$&$\begin{tabular}{c}
2.81-5.62 \\
3.03-6.06 \\
1.81-3.62 \\
1.97-3.93 \\
\end{tabular}$&$\begin{tabular}{c}
7.65-15.5 \\
8.25-16.7 \\
4.95-9.90 \\
5.40-10.8 \\
\end{tabular}$ \\
\hline
$Total$&$3-6$&$\begin{tabular}{c}
1a  \\
1a  \\
1b  \\
1b  \\
\end{tabular}$&$\begin{tabular}{c}
$30^{\circ}$ \\
$35^{\circ}$ \\
$30^{\circ}$ \\
$35^{\circ}$ \\
\end{tabular}$&
$\begin{tabular}{c}
1 \\
1 \\
1 \\
1 \\
\end{tabular}$&$\begin{tabular}{c}
1.65-3.30  \\
2.10-3.90  \\
1.50-2.85  \\
1.80-3.60  \\
\end{tabular}$&$\begin{tabular}{c}
3.75-7.50  \\
4.35-9.00  \\
3.30-6.45  \\
4.05-8.10 \\
\end{tabular}$&$\begin{tabular}{c}
4.36-8.72  \\
5.19-10.4  \\
3.75-7.50  \\
4.72-9.44 \\
\end{tabular}$&$\begin{tabular}{c}
11.9-23.9  \\
14.1-28.5  \\
10.2-20.6 \\
12.9-25.9 \\
\end{tabular}$ \\
\hline
\end{tabular}
\label{Nev-T}
\end{table*}

The IMF-1 is not very realistic since it is based upon counts of M
class stars in the Galactic disk (Gould et al. 1997) and should
thus be taken just as an illustration. Moreover, as can be
inferred from Table \ref{Nev-T}, for the IMF-1  mass distribution
the number of events is less than for the the more realistic IMF-2
and IMF-3 cases. The IMF-4 is also not very realistic, leading
already for a rather small dispersion velocity to a conspicuous
number of events. However, as we will see below, the resulting
average event duration for the IMF-4 is definitely not in
agreement with the observed value.

For the IMF-2 and the IMF-3 we see that even by considering the
more extreme values for the various parameters we get at most
about 10 events, which is still less then the ones reported by the
MACHO group although not very much. However, this implies
to adopt rather unlikely values for the various LMC parameters.
For the preferred and more realistic values, leading to an
optical depth of $\tau \simeq 2 - 3 \times 10^{-8}$, we expect about
3 to 5 events due to self-lensing (Table \ref{Nev-T}).
\vspace{0.5 cm}
\item[$\bullet$] Model 2.
For the model 2, on the ground of the results from the previous
section, we concentrate only on the more realistic models for
self-lensing by using IMF-2 and IMF-3. The results for the
different source/lens geometries are shown in Table \ref{Nev-2}
for model $2a$ and $2b$, where we fixed  again the detection
efficiency to 0.5 and the velocity dispersion to 30 km/sec. We see
that also these models lead to values which are similar to the
ones of model 1 and thus are also not able to explain all the
observed MACHO events. For the model 2 we expect again rougly $3$
events due to self-lensing when using the lower value for the LMC
mass. Thus it seems that this conclusion is rather robust and does
not depend much on the detailed form of the bar and the disk.
\end{itemize}
\begin{table*}
\centering \caption{Model 2: estimates of the total number of
microlensing events towards LMC for the MACHO observation time
computed in the case of self-lensing and for different values of
the velocity dispersion, LMC mass and efficiency. We used
$\varepsilon=0.5$ for the detection efficiency and
$\sigma=30\,\mathrm{km \, sec^{-1}}$ for the velocity dispersion.
Two Salpeter-type IMFs were used (see text for the values of the
parameters).}
\begin{tabular}[h]{cccccc}
\hline \hline $Source/Lens$&$LMC \, mass$&
$Model$&$Relative$&$N_{\mathrm{ev}}$&$N_{\mathrm{ev}}$\\
$Geometry$&$(\times10^{9}\mathrm{M_{\sun}})$&$$&$Weight$&$(\mathrm{IMF-2})$&
$(\mathrm{IMF-3})$\\
\hline $Disk/disk$&$\begin{tabular}{c}
3 - 6  \\
3 - 5  \\
\end{tabular}$&$\begin{tabular}{c}
2a  \\
2b  \\
\end{tabular}$&$\begin{tabular}{c}
0.50 \\
0.20 \\
\end{tabular}$&$\begin{tabular}{c}
2.38 - 4.77 \\
1.49 - 2.48\\
\end{tabular}$&$\begin{tabular}{c}
2.76 - 5.53 \\
1.73 - 2.88 \\
\end{tabular}$ \\
\hline
$Disk/bar$&$\begin{tabular}{c}
3 - 6  \\
3 - 5  \\
\end{tabular}$&$\begin{tabular}{c}
2a  \\
2b  \\
\end{tabular}$&$\begin{tabular}{c}
0.50 \\
0.29 \\
\end{tabular}$&$\begin{tabular}{c}
0.58 - 1.17 \\
1.46 - 2.43 \\
\end{tabular}$&$\begin{tabular}{c}
0.68 - 1.35 \\
1.69 - 2.82 \\
\end{tabular}$ \\
\hline
$Bar/disk$&$\begin{tabular}{c}
3 - 6  \\
3 - 5  \\
\end{tabular}$&$\begin{tabular}{c}
2a  \\
2b  \\
\end{tabular}$&$\begin{tabular}{c}
0.50 \\
0.80 \\
\end{tabular}$&$\begin{tabular}{c}
2.11 - 4.23 \\
1.32 - 2.21 \\
\end{tabular}$&$\begin{tabular}{c}
2.45 - 4.91 \\
1.53 - 2.56 \\
\end{tabular}$ \\
\hline
$Bar/bar$&$\begin{tabular}{c}
3 - 6  \\
3 - 5  \\
\end{tabular}$&$\begin{tabular}{c}
2a  \\
2b  \\
\end{tabular}$&$\begin{tabular}{c}
0.50 \\
0.80 \\
\end{tabular}$&$\begin{tabular}{c}
0.61 - 1.22 \\
1.52 - 2.53 \\
\end{tabular}$&$\begin{tabular}{c}
0.70 - 1.41 \\
1.76 - 2.93 \\
\end{tabular}$ \\
\hline
$Total$&$\begin{tabular}{c}
3 - 6  \\
3 - 5  \\
\end{tabular}$&$\begin{tabular}{c}
2a  \\
2b  \\
\end{tabular}$&
$\begin{tabular}{c}
1 \\
1 \\
\end{tabular}$&$\begin{tabular}{c}
2.84 - 5.70 \\
2.86 - 4.77 \\
\end{tabular}$&$\begin{tabular}{c}
3.30 - 6.60 \\
3.32 - 5.53 \\
\end{tabular}$ \\
\hline
\end{tabular}
\label{Nev-2}
\end{table*}

Similarly, we can estimate the number of self-lensing events we
expect for the EROS2 data for a disk/disk source lens geometry. In
this case, using the IMF-2 and an efficiency equal to about 0.12
(see Milsztajn \& Lasserre 2001), the estimated number of events,
using the model $1a$, is $0.16-0.32$ for
$\sigma=30\,\mathrm{km/sec}$ and for a LMC disk mass in between
$2.4-4.8\times10^{9}\mathrm{M_{\sun}}$. For the IMF-3 the
corresponding number of events is $0.19-0.37$. If instead we use
the model $2a$, we expect for the above parameters a number of
events equal to $0.23-0.46$ using the IMF-2, and $0.27-0.53$ using
the IMF-3. Finally, we expect $0.14-0.24$ and $0.17-0.28$ for the
model 2b and for IMF-2 and IMF-3, respectively. Even if the small
number of events found by EROS2 does not allow to draw a
definitive conclusion, it seems unlikely that the EROS2 events are
all due to self-lensing.

We computed also the expected number of events due to MACHOs in
the LMC halo. The lens number density is given by
\begin{equation}
\label{lmchalo}
\frac{dn}{d\mu}=\left(1+\frac{R^{2}}{a^{2}}\right)^{-1}\frac{dn_{0}}{d\mu},
\end{equation}
where for $dn_0/d\mu$ we assume a delta function, and fix $a$
equal to 2 kpc (Gyuk et al. 2000) (see Table \ref{G-Nev-T}). Using
the MACHO collaboration parameters and $\mu=0.5$ we expect roughly
5 events under the assumption of a 100\% MACHO LMC halo and
assuming that the rate is uniformly equal to the one for the line
of sight going through the center of the LMC, this last assumption
obviously leads to an upper limit. Clearly, since the halo mass
fraction in form of MACHOs in the halo of our Galaxy is at most
20\% (Alcock et al. 2000a), we do not expect this fraction to be
higher in the LMC halo. Thus, realistically we expect at most
about 1 event coming from this component. A similar conclusion
holds for the EROS2 data.
%
%
\subsection{Average event duration}
\label{self-lensing_times}
Using the Eq. (\ref{average-event-duration}) we get estimates of
the average event duration $<T>$ for different source/lens
geometries and IMFs. The results, averaged on 27 MACHO field, are
shown in Table \ref{T} for model 1 and in Table \ref{T2} for model
2. The velocity dispersion is in both cases assumed to be
$30\,\mathrm{km/sec}$.
\begin{table*}
\centering \caption{Model 1: theoretical estimates of the LMC
self-lensing event durations, $<T>$, averaged over 27 MACHO
fields. We used $\sigma=30\,\mathrm{km \, sec^{-1}}$ for the
velocity dispersion. Different source/lens geometries and four
Salpeter-type IMFs were used (For the parameters see text). The
durations are given in days.}
\begin{tabular}[h]{ccccccc}
\hline \hline $Source/Lens$&$Model$&$LMC$&$<T>$ &$<T>$&$<T>$&$<T>$\\
$Geometry$&$$&$inclination$&$(IMF-1)$&$(IMF-2)$&$(IMF-3)$&$(IMF-4)$ \\
\hline $Disk/disk$&$\begin{tabular}{c}
1a  \\
1a  \\
1b  \\
1b  \\
\end{tabular}$&$\begin{tabular}{c}
$30^{\circ}$ \\
$35^{\circ}$ \\
$30^{\circ}$ \\
$35^{\circ}$ \\
\end{tabular}$&$\begin{tabular}{c}
67.3 \\
68.8 \\
67.3 \\
68.8 \\
\end{tabular}$&$\begin{tabular}{c}
49.5  \\
50.7  \\
49.5  \\
50.7  \\
\end{tabular}$&$\begin{tabular}{c}
42.9  \\
43.9  \\
42.9  \\
43.9  \\
\end{tabular}$&$\begin{tabular}{c}
19.9  \\
20.3  \\
19.9  \\
20.3  \\
\end{tabular}$ \\
\hline
$Disk/bar$&$\begin{tabular}{c}
1a  \\
1a  \\
1b  \\
1b  \\
\end{tabular}$&$\begin{tabular}{c}
$30^{\circ}$ \\
$35^{\circ}$ \\
$30^{\circ}$ \\
$35^{\circ}$ \\
\end{tabular}$&$\begin{tabular}{c}
59.6 \\
60.7 \\
63.4 \\
64.3 \\
\end{tabular}$&$\begin{tabular}{c}
43.9 \\
44.7 \\
46.7 \\
47.3 \\
\end{tabular}$&$\begin{tabular}{c}
38.0 \\
38.7 \\
40.4 \\
41.0 \\
\end{tabular}$&$\begin{tabular}{c}
17.6 \\
17.9 \\
18.7 \\
19.0 \\
\end{tabular}$ \\
\hline
$Bar/disk$&$\begin{tabular}{c}
1a  \\
1a  \\
1b  \\
1b  \\
\end{tabular}$&$\begin{tabular}{c}
$30^{\circ}$ \\
$35^{\circ}$ \\
$30^{\circ}$ \\
$35^{\circ}$ \\
\end{tabular}$&$\begin{tabular}{c}
75.1 \\
80.4 \\
82.3 \\
87.1 \\
\end{tabular}$&$\begin{tabular}{c}
55.3 \\
59.2 \\
60.6 \\
64.1 \\
\end{tabular}$&$\begin{tabular}{c}
47.9 \\
51.3 \\
52.5 \\
55.5 \\
\end{tabular}$&$\begin{tabular}{c}
22.2 \\
23.7 \\
24.3 \\
25.7 \\
\end{tabular}$ \\
\hline
$Bar/bar$&$\begin{tabular}{c}
1a  \\
1a  \\
1b  \\
1b  \\
\end{tabular}$&$\begin{tabular}{c}
$30^{\circ}$ \\
$35^{\circ}$ \\
$30^{\circ}$ \\
$35^{\circ}$ \\
\end{tabular}$&$\begin{tabular}{c}
54.3 \\
54.3 \\
65.4 \\
65.3 \\
\end{tabular}$&$\begin{tabular}{c}
40.0 \\
40.0 \\
48.1 \\
48.1 \\
\end{tabular}$&$\begin{tabular}{c}
34.6 \\
34.6 \\
41.7 \\
41.7 \\
\end{tabular}$&$\begin{tabular}{c}
16.0 \\
16.0 \\
19.3 \\
19.3 \\
\end{tabular}$ \\
\hline
$Average$&$\begin{tabular}{c}
1a  \\
1a  \\
1b  \\
1b  \\
\end{tabular}$&$\begin{tabular}{c}
$30^{\circ}$ \\
$35^{\circ}$ \\
$30^{\circ}$ \\
$35^{\circ}$ \\
\end{tabular}$&$\begin{tabular}{c}
64.1  \\
66.1  \\
69.6 \\
71.4  \\
\end{tabular}$&$\begin{tabular}{c}
47.2  \\
48.7  \\
51.2 \\
52.6 \\
\end{tabular}$&$\begin{tabular}{c}
40.9 \\
42.1  \\
44.4  \\
45.5 \\
\end{tabular}$&$\begin{tabular}{c}
18.9  \\
19.5 \\
20.6 \\
21.1 \\
\end{tabular}$ \\
\hline
\end{tabular}
\label{T}
\end{table*}

\begin{figure}
 \resizebox{\hsize}{!}{\includegraphics{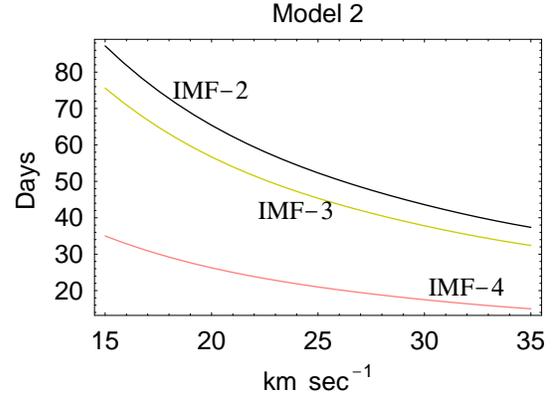}}
 \caption{Model 2: variation of the total self-lensing event duration
 as a function of the velocity
 dispersion. These curves are obtained by using a LMC disk
 inclination of $35^{\circ}$. Three different IMFs were used
(for the parameters see text).}
 \label{durations-2}
\end{figure}

\begin{itemize}
\item[$\bullet$]Model 1.
In Fig. \ref{Self-Tempi} we report the average event duration
obtained using the four considered IMFs and different values of
the velocity dispersions.
\begin{figure}
 \resizebox{\hsize}{!}{\includegraphics{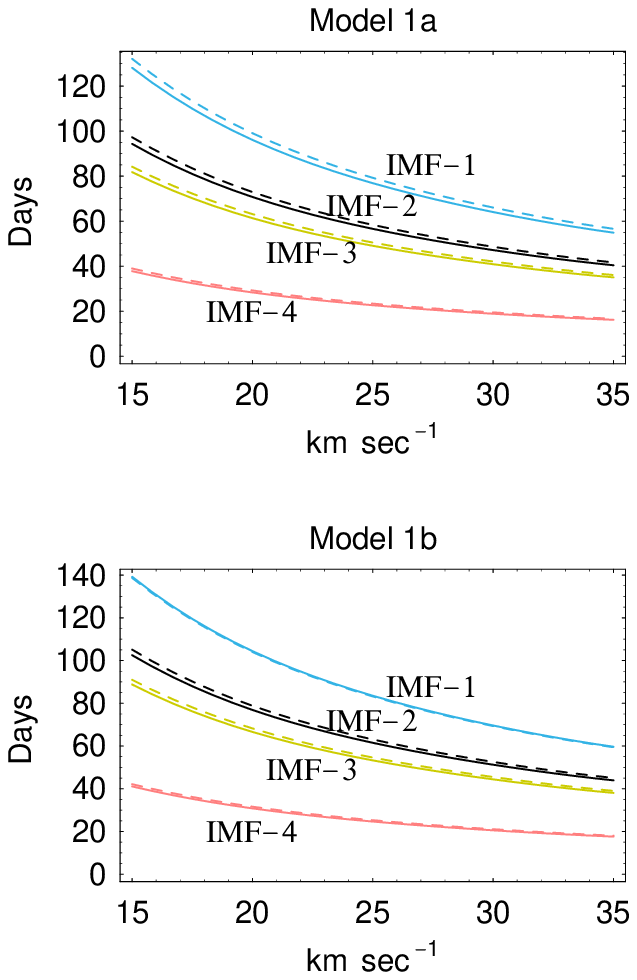}}
 \caption{Model 1: variation of the total self-lensing event
 durations as a function of the velocity
 dispersion. The solid lines and the dashed lines are obtained
 by using a LMC disk inclination of $30^{\circ}$ and $35^{\circ}$,
 respectively. Four different IMFs were used (for the parameters
 see text). Two models ($1a$ and $1b$) are shown separately.}
 \label{Self-Tempi}
\end{figure}
We recall that the IMF-1 was deduced by dwarf star counts in the
Milky Way disk, so we must be cautious when applying it to a
population of another galaxy like LMC.

The results in Fig. \ref{Self-Tempi} can not be directly compared
with the observed MACHO event durations as summarized in Table
\ref{MACHO-times}, because among them there are two events, LMC-9
(binary event) and LMC-22 (exotic event), which are not compatible
with our method of analysis and there is one more event, LMC-5
(disk event) which is not a self-lensing event.
\begin{table}
\centering \caption{EROS and MACHO (blended and unblended) average
event durations. The quantity $\hat{t}_{st}(A,B)$ is the average
event timescale, according to the two different criteria used in
the efficiency, for events in the MACHO Monte Carlo calculations
that have been detected with an unblended fit timescale of
$\hat{t}$.}
\begin{tabular}{cccc}
\hline \hline
$Team$&$Set$&$N_{\mathrm{ev}}$&$<T>$\\
\hline
$\mbox{EROS}$&$\hat{t}$&$5$&$33\,\mathrm{days}$ \\
$\mbox{MACHO}$&$\hat{t}$&$17$&$44\,\mathrm{days}$ \\
$\mbox{MACHO}$&$\hat{t}_{st}(A)$&$13$&$47\,\mathrm{days}$ \\
$\mbox{MACHO}$&$\hat{t}_{st}(B)$&$17$&$56\,\mathrm{days}$ \\
\hline
\end{tabular}
\label{MACHO-times}
\end{table}

So, we must exclude these three events from the MACHO blended
average event durations, set $\hat{t}$, before comparing with our
results. The remaining $14$ events give an average duration of $37
\, \mathrm{days}$. Moreover, as mentioned above (see also Fig.
\ref{istogramma}) all relevant events have a duration in between
18 and 52 days. Comparing this with our results (Fig.
\ref{Self-Tempi}) and taking also into account that not all events
are possibly due to self-lensing, clearly leads already to
restrictions on the IMF and in particular on the velocity
dispersion values: the velocity dispersion is more likely to be in
the range of $30-35\,\mathrm{km/sec}$, that suggests a
predominance of lenses of old population (Hughes et al. 1991) and
IMF-2 or IMF-3 are favored, since IMF-1 leads to too high
durations. On the other hand an IMF-4 model would require a low
velocity dispersion (below $20\,\mathrm{km/sec}$), otherwise the
average duration gets substantially below the observed value.
However, for very low mass objects one would rather expect a
higher velocity dispersion.

\vspace{0.5 cm}
\item[$\bullet$] Model 2.

For this model we consider three IMFs: IMF-2, IMF-3 and IMF-4. The
results, reported in Table \ref{T2} and in Fig. \ref{durations-2},
show that the mean durations are slightly lower than that of model
1, nonetheless the previous conclusions hold.
\end{itemize}
\begin{table}
\centering \caption{Model 2: theoretical estimates of the LMC
self-lensing durations of events, $<T>$, averaged over 27 MACHO
fields. We used $\sigma=30\,\mathrm{km \, sec^{-1}}$ for the
velocity dispersion. Various source/lens geometries and three
Salpeter-type IMFs were used (for the parameters see text). The
durations are given in days. (We do not distinguish between model
2a and 2b, since both give the same results.)}
\begin{tabular}{ccccc}
\hline \hline $Source/Lens$&$<T>$&$<T>$&$<T>$  \\
$Geometry$&$(IMF-2)$&$(IMF-3)$&$(IMF-4)$ \\
\hline
$Disk/disk$&$52.6$&$45.6$&$21.1$ \\
$Disk/bar$&$46.3$&$40.1$&$18.5$ \\
$Bar/disk$&$33.8$&$29.3$&$13.6$ \\
$Bar/bar$&$41.6$&$36.0$&$16.7$ \\
$Average$&$43.6$&$37.8$&$17.5$ \\
\hline
\end{tabular}
\label{T2}
\end{table}
We arrive at the same conclusion also if we compare our results
with the EROS mean event duration (see Table \ref{MACHO-times})
under the rather unplausible hypothesis that they are all due to
self-lensing.

For the IMF-4 mass distribution we see from Fig. \ref{Self-Tempi}
that the duration of events decreases slowly by increasing
the velocity dispersion of the lenses and in order to explain
the observed mean duration of the events requires a low ($\sigma <
20 \, \mathrm{km/sec}$) velocity dispersion. For what we saw in
the previous section on the IMF-4 model this would then also imply
that all the detected events are due to self-lensing. However,
such a low required dispersion velocity is rather unplausible
considering also the small mass of the lenses. Thus the IMF-4
model is certainly not very realistic.

Using the number density given by Eq. (\ref{lmchalo}), we
estimated the event duration also for lenses in the LMC halo, see
Table \ref{G-Nev-T}. We expect a duration in between $128
\sqrt{\mu}$ and $181 \sqrt{\mu}$ days according if we take into
account the motion of the source as in the former case or not.
%

\subsection{Self-lensing mean mass}
\label{Self-lensing-mean-mass}
With the mass moment method we are able to get an estimate of the
lens mean mass $<M>$, as given by Eq. (\ref{mean-mass}). We notice
that to get this way the mean mass we do clearly not need to
assume a particular IMF distribution. For $<T^{1}>$ and
$<T^{-1}>$, we used the Einstein radius crossing times measured by
the MACHO collaboration for the 17 events that they detected
(Alcock et al. 2000a). Some of them were excluded in our LMC
self-lensing analysis: the Galactic disk event LMC-5, the binary
event LMC-9, and the exotic event LMC-22 (probably the source is a
distant active galaxy (Bennett 2001)). With this exclusion,
the events LMC-1 and LMC-15 have the shortest duration, while the
events LMC-7, LMC-13 and LMC-14 have the longest duration.
Moreover, LMC-14 is a self-lensing event (Alcock et al. 2001b). In
Section \ref{number of events} we saw that the expected number of
self-lensing events is at least 3. Assuming thus that among the
observed events (not including the 3 excluded ones as explained
above, one of which being also probably a self-lensing) just 3 are
due to self-lensing, we can estimate a minimum and a maximum mean
mass using the following 3 events: LMC-1, LMC-14, LMC-15 (mean
duration: 28.5 days) for $<M_{\mathrm{min}}>$, and LMC-7, LMC-13,
LMC-14 (mean duration: 50.5 days) for $<M_{\mathrm{max}}>$.

\begin{figure}
 \resizebox{\hsize}{!}{\includegraphics{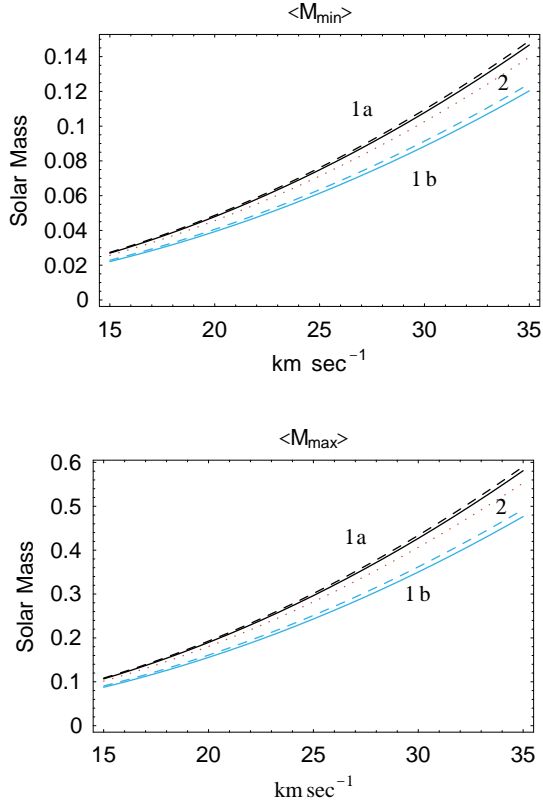}}
 \caption{Total mean minimal and maximal mass of the lenses
in solar mass unity as a function of the velocity dispersion
(km/sec) for models $1a$, $1b$ and $2$ (dotted line). For the
models $1a$ and $1b$, the solid lines and the dashed lines are
obtained by using a LMC disk inclination of $30^{\circ}$ and
$35^{\circ}$, respectively.}
 \label{Mean-mass}
\end{figure}

In Fig. \ref{Mean-mass} we report the results for different values
of the velocity dispersion obtained using models $1a$, $1b$ and
$2$ (we do not distinguish between
model $2a$ and $2b$, since both give the same results).
Any other combination in the choice of the events (or in
particular by assuming that more than 3 events are due to
self-lensing) would lead to a higher value for
$<M_{\mathrm{min}}>$ and to a lower value for
$<M_{\mathrm{max}}>$. Thus the values given here have to be
considered as lower and upper limit, respectively. These results
have a slight dependence from the LMC disk inclination, as it
appears from Fig. \ref{Mean-mass}, which shows also the variation
of the mean mass for a disk inclination between
$30^{\circ}-35^{\circ}$.

We see that for our preferred values for the LMC mass
and for $\sigma \simeq
30-35\,\mathrm{km\,sec^{-1}}$, we get a mass in the range $\sim
0.1 - 0.6\,\mathrm{M_{\sun}}$, which is consistent with the
various assumptions we made. This result suggests that the lenses
are low mass stars, which is certainly not surprising This can
also be seen as a consistency check on the modelling of the LMC we
adopted.

\section{Galactic and halo lensing populations}
In this section we turn to the case that some of the lenses are
located in the Milky Way or in its halo. A Galactic model with
three main components fits well all the dynamical observations.
These components are the disk, the spheroid and the halo. The
halo, composed essentially of dark matter, contributes for most
part of the total mass of the Galaxy and should give the higher
contribution to the microlensing surveys. However, since at least
one event (LMC-5) is due to a Galactic disk lens (Alcock et al.
2001c), we will examine two other important Galactic components:
disk (thin and thick) and spheroid. In Tables \ref{G-Tau} and
\ref{G-Nev-T} we summarize the estimates of the microlensing
quantities related to these components. The thin and thick disks
are considered as separate components.

There is no general agreement on the details of the structure of
our Galaxy and in particular on the parameters of its thick disk
and spheroid. For the disk we vary the structural parameters in a
wide range, while for the spheroid we refer to Giudice et al.
(1994), where the modelling of this component is amply discussed.
Moreover, we refer to Carrol \& Ostlie (1996), Roulet \& Mollerach
(1997) and Sparke \& Gallagher (2000) for most of the parameters
used to describe the Galactic components.
\begin{table*}
\centering \caption{Theoretical estimates of the optical depth for
stars in the LMC due to different lensing populations: thin disk,
thick disk, spheroid (stellar halo), MW halo and LMC halo.}
\begin{tabular}[h]{cccc}
\hline \hline $Lensing \,
component$&$\Sigma_{0}\,(\mathrm{M_{\sun}pc^{-2}})$&
$\rho_{0}\,(\mathrm{M_{\sun}pc^{-3}})$&$\tau$\\
\hline
$Thin \, Disk$&$(30-50)$&$-$&$(0.92-1.54)\times10^{-8}$ \\
$Thick \, Disk$&$(35-45)$&$-$&$(3.09-3.97)\times10^{-8}$ \\
$Spheroid$&$-$&$0.0015\times10^{-3}$&$4.29\times10^{-8}$ \\
$MW\, halo$&$-$&$0.0079\times10^{-3}$&$4.83\times10^{-7}$ \\
$LMC \, halo$&$-$&$0.0223\times10^{-3}$&$8.36\times10^{-8}$ \\
\hline
\end{tabular}
\label{G-Tau}
\end{table*}
\begin{table*}
\centering \caption{Theoretical estimates of the number of events
and event mean duration for LMC stars by different lensing
populations: thin disk, thick disk, spheroid (stellar halo), MW
halo and LMC halo. A $\delta$-function was used to describe the
mass distribution of the lenses in the halos. Instead, two IMFs
with different slopes were used to describe the other Galactic
components. Different values for the dispersion velocity,
$\sigma$, were used. $\varepsilon$ is the MACHO detection
efficiency and $\mu$ is the lens mass in solar mass units.}
\begin{tabular}[h]{cccccc}
\hline \hline $Component$&$\sigma$&$dn_{0}/d\mu$&$\mbox{MACHO} \,
N_{\mathrm{ev}}$&
$\mbox{EROS} \,N_{\mathrm{ev}}$&$<T>$\\
\hline $Thin  \,
Disk$&$(20-30)\,\mathrm{km\,sec^{-1}}$&$\begin{tabular}{c}
IMF-1 \\
IMF-2 \\
\end{tabular}$&$\begin{tabular}{c}
$(0.3-0.4)$ $\varepsilon$ \\
$(0.9-1.4)$ $\varepsilon$
\end{tabular}$&$\begin{tabular}{c}
$0.04-0.05$ \\
$0.12-0.18$ \\
\end{tabular}$&$\begin{tabular}{c}
$(107-71)$ \,$\mathrm{days}$ \\
$(78-52)$ \, $\mathrm{days}$
\end{tabular}$\\
\hline $Thick  \,
Disk$&$(35-60)\mathrm{km\,sec^{-1}}$&$\begin{tabular}{c}
IMF-1 \\
IMF-2 \\
\end{tabular}$&$\begin{tabular}{c}
$(1.5-2.5)$ $\varepsilon$ \\
$(3.3-5.7)$ $\varepsilon$ \\
\end{tabular}$&
$\begin{tabular}{c}
$0.2-0.3$  \\
$0.5-0.8$  \\
\end{tabular}$&$\begin{tabular}{c}
$(103-60)$\, $\mathrm{days}$\\
$(76-44)$\, $\mathrm{days}$\\
\end{tabular}$\\
\hline
$Spheroid$&$(90-120)\,\mathrm{km\,sec^{-1}}$&$\begin{tabular}{c}
IMF-1 \\
IMF-2 \\
\end{tabular}$&$\begin{tabular}{c}
$(2.8-3.8)$ $\varepsilon$ \\
$(6.5-8.7)$ $\varepsilon$ \\
\end{tabular}$&$\begin{tabular}{c}
$0.4-0.5$  \\
$0.9-1.2$  \\
\end{tabular}$&$\begin{tabular}{c}
$(69-52)$\, $\mathrm{days}$ \\
$(51-38)$\, $\mathrm{days}$ \\
\end{tabular}$\\
\hline $MW\,halo$&$210\,\mathrm{km\,sec^{-1}}$&$\delta$&$107.7\,
\varepsilon\,\mu^{-1/2}$&$14.6\,\mu^{-1/2}$&
$71\,\sqrt{\mu}\,\mathrm{days}$ \\
\hline
$LMC\,halo$&$\mathrm{50\,km\,sec^{-1}}$&$\delta$&$7.3\,\varepsilon
\mu^{-1/2}$&$1.0\,\mu^{-1/2}$&
$(128-181)\,\sqrt{\mu}\,\mathrm{days}$ \\
\hline
\end{tabular}
\label{G-Nev-T}
\end{table*}
%

\subsection{Lenses in the Galactic Thick Disk}
We start our analysis  on galactic components with the hypothesis
that some lenses are associated with the oldest sub-population of
the Galactic disk. We consider that these lenses, a fraction of
which might be high velocity white dwarfs (Reid et al. 2001), are
confined within the so-called {\it thick disk}. In this case, we
used a lens number density distribution per unit mass $dn/d\mu$
given by
\begin{eqnarray}\label{thick-disk}
&&\frac{dn}{d\mu}=H_{\mathrm{TD}}\frac{dn_{0}}{d\mu}=\nonumber \\ &=&
\exp\left[-\frac
{\sqrt{R_{0}^{2}+D_{\mathrm{ol}}^{2}\cos^2{b}-2R_{0}D_{\mathrm{ol}}\cos{b}\cos{l}}}
{R_{\mathrm{TD}}}
\right]\times\\ &\times &
\exp\left[\frac{R_{0}}{R_{\mathrm{TD}}}\right]
\exp\left[-\frac
{|D_{\mathrm{ol}}\sin{b}|}
{z_{\mathrm{TD}}}
\right]\frac{dn_{0}}{d\mu}\nonumber,
\end{eqnarray}
where $R_{0}$ is the Earth distance from the center of the Galaxy,
for which we use the IAU recommended value of $8.5$ kpc. The
common value for the thick disk scale length $R_{\mathrm{TD}}$ is
$3.5 \,\mathrm{kpc}$ (Bahcall 1986), even if its estimate varies
in a wide range. Here, we will vary $R_{\mathrm{TD}}$ in between 2
and 5 kpc. The thick disk has a larger scale height as compared to
the thin disk. Its value is generally estimated to be in the range
$1<z_{\mathrm{TD}}<1.5$ kpc. For the local column density we use
$\Sigma_{0}^{\mathrm{TD}}=35-45\, \, \mathrm{M_{\sun} \, pc^{-2}}$
(Derue et al. 2001; Roulet \& Mollerach 1997). All the estimates
of microlensing quantities for $R_{\mathrm{TD}}=3.5$ kpc, and
$z_{\mathrm{TD}}=1$ (Bahcall 1986), are reported in Tables
\ref{G-Tau} and \ref{G-Nev-T}.

\begin{figure*}
 \resizebox{\hsize}{!}{\includegraphics{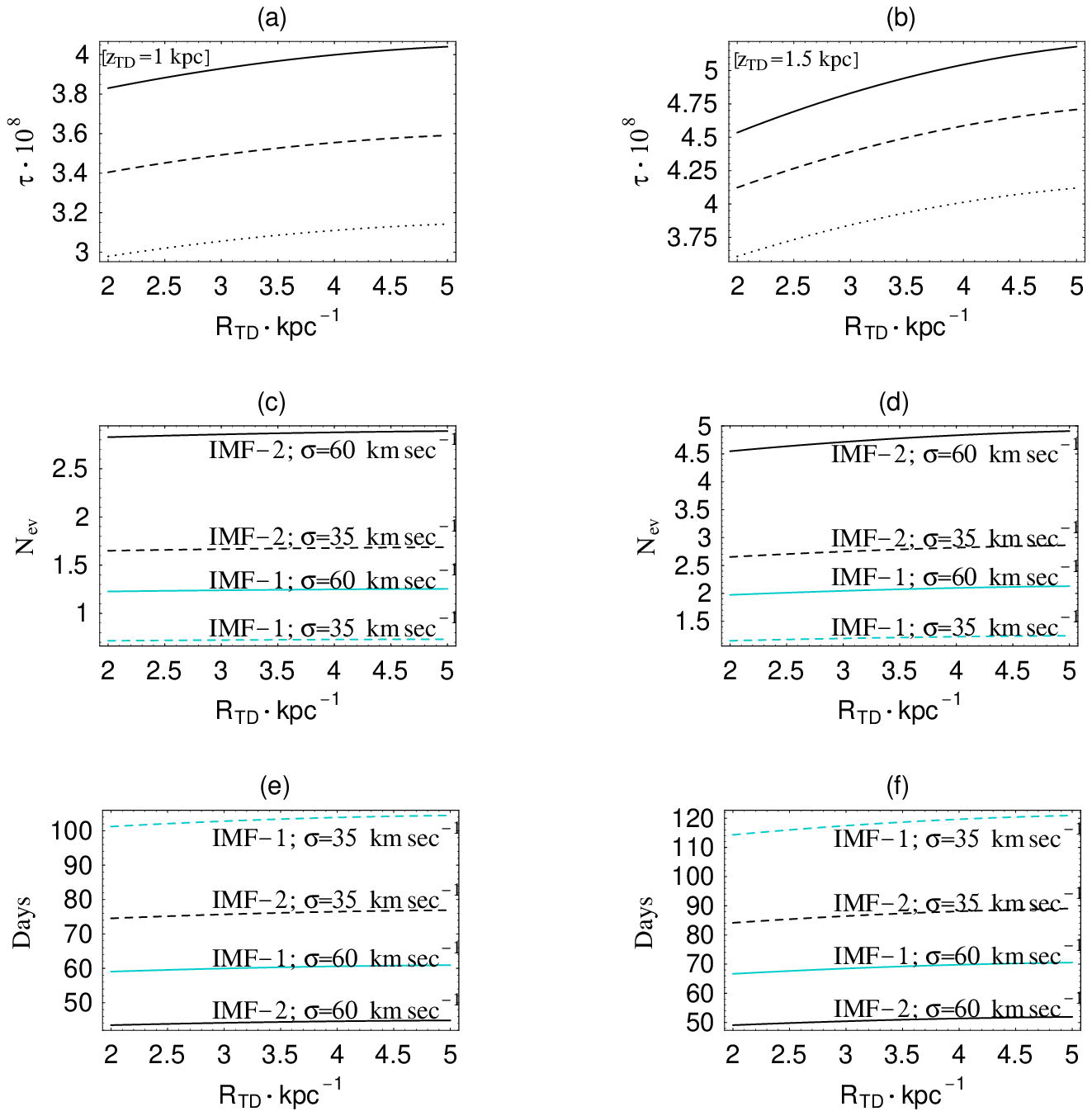}}
 \caption{Galactic thick disk microlensing quantities as a function of the scale length,
 $R_{\mathrm{TD}}$, for different local column densities, IMFs, dispersion velocities,
 and for the scale height, $z_{\mathrm{TD}}$, equal to 1 kpc (a, c, e) and
 1.5 kpc (b, d, f). MACHO collaboration experimental values were used.
 Here we take
 the detection efficiency $\varepsilon$ equal to 0.5. (a, b) Optical depth:
 the solid line was estimated for
  $\Sigma_{0}^{\mathrm{TD}}=45\, \mathrm{M}_{\sun}\, \mathrm{pc}^{-2}$, the
  dashed line for $\Sigma_{0}^{\mathrm{TD}}=40\, \mathrm{M}_{\sun}\, \mathrm{pc}^{-2}$,
  and the dotted line for $\Sigma_{0}^{\mathrm{TD}}=35\, \mathrm{M}_{\sun}\,
  \mathrm{pc}^{-2}$. (c, d) Number of events. (e, f) Mean duration of
  events.}
 \label{thickdisk}
\end{figure*}

In Fig. \ref{thickdisk} we plotted the variation of the optical
depth as a function of $R_{\mathrm{TD}}$ for different values of
$\Sigma_{0}^{\mathrm{TD}}$ and with values for $z_{\mathrm{TD}}$
equal to 1 kpc (Fig. \ref{thickdisk}a) and to 1.5 kpc (Fig.
\ref{thickdisk}b), respectively.
From these figures one sees that
the optical depth varies little with the scale length
$R_{\mathrm{TD}}$. Instead, there is a more clear dependence of
$\tau$ on the value of the scale height $z_{\mathrm{TD}}$, besides
also the density. This is true also for the number and the mean
duration of events, Fig. \ref{thickdisk}c-d and Fig.
\ref{thickdisk}e-f respectively, even if in these cases the
parameters on which one has to vary are connected to the IMF and
the dispersion velocity of the lenses.

For $R_{\mathrm{TD}}=3.5$ kpc, $z_{\mathrm{TD}}=1$, and
$\Sigma_{0}^{\mathrm{TD}}=35-45 \, \mathrm{M_{\sun} \, pc^{-2}}$,
we found that the optical depth towards the LMC is
$\tau=3.09-3.97\times 10^{-8}$ (Table \ref{G-Tau}), that is higher
than the estimated white dwarf population optical depth
$\tau\approx 0.4\times 10^{-8}$ (Koopmans \& Blandford 2001).

Following the discussion in Sect. $3.3$, for the mass distribution
we used two Salpeter type IMFs: IMF-1 with $\alpha=0.56$ for
$0.1<\mu<0.59$ and $\alpha=2.2$ for $\mu>0.59$; IMF-2 with
$\alpha=2.35$. For MACHO experimental numbers, in Fig.
\ref{thickdisk} we report the variation of the number of events as
a function of $R_{\mathrm{TD}}$ for different values of $\sigma$
and adopting $\varepsilon=0.5$ (Alcock et al. 2000a),
$z_{\mathrm{TD}}$ equal to 1 kpc (Fig. \ref{thickdisk}c) and to
1.5 kpc (Fig. \ref{thickdisk}d).

Here and in the next section, clearly the use of IMF-1 is
favoured, because it was obtained from Galactic disk observations,
as we already mentioned in Sect. \ref{number of events}. In this
case, joining thin and thick disk, we expect $1-3$ events for
$R_{\mathrm{TD}}=3.5$ kpc, $z_{\mathrm{TD}}=1$, and for the
velocity dispersion in the range $\sigma_{\mathrm{TD}}=35-60 \,
\mathrm{km\, sec^{-1}}$ (Derue et al. 2001; Carrol \& Ostlie
1996).

At last, we estimated the mean duration of the events as a
function of $R_{\mathrm{TD}}$ for different values of $\sigma$ and
adopting $z_{\mathrm{TD}}$ equal to 1 kpc (Fig. \ref{thickdisk}e)
and to 1.5 kpc (Fig. \ref{thickdisk}f). It seems that a disk event
was really observed, LMC-5 (Alcock et al. 2001c). Its duration,
$37.8\, \mathrm{days}$, is not in good agreement with our
predictions (Table \ref{G-Nev-T}), however one can not draw any
conclusion on just one event. Moreover, we assumed a minimum mass
in the IMF-1 of 0.1 $\mathrm{M}_{\sun}$, whereas from the
observation it turns out that the lens mass could well be below
that limit, explaining thus its shorter duration.
%

\subsection{Lenses in the Galactic Thin Disk}
We consider now the possibility to have lenses in the thin disk.
This component, contrary to the thick disk that is composed of low
metallicity stars, is made  of relatively young metal-rich stars
with a small velocity dispersion $\sigma_{\mathrm{td}} \simeq
20-30 \, \mathrm{km \, sec^{-1}}$ (Roulet \& Mollerach 1997). We
modelled it as in the previous case, see Eq. (\ref{thick-disk}),
but with a smaller scale height $z_{\mathrm{td}}=0.1-0.35 \,
\mathrm{kpc}$ (Roulet \& Mollerach 1997) and with a local column
density $\Sigma_{0}^{\mathrm{td}}=30-50 \, \mathrm{M_{\sun}\,
pc^{-2}}$ (Roulet \& Mollerach 1997; Derue et al 2001).

For the thick disk we saw that the microlensing quantities do not
vary much as a function of the scale length, but there were
variations with the scale height. For the thin disk the scale
length is know with only a small range of uncertainty and,
therefore, the modifications on our theoretical estimates will be
negligible. Here we use $z_{\mathrm{td}}=0.3 \, \mathrm{kpc}$
(Bahcall 1986), and the optical depth turns out to be
$\tau=0.9-1.5 \times 10^{-8}$, see Table \ref{G-Tau}.

Using again two different IMFs as in the previous case, we
estimated the number of events and their mean duration. The
results are reported in Table \ref{G-Nev-T}. We see that the
expected average event duration for both the thick and thin disk
components is above 50 days when considering the more realistic
IMF-1, so that it is rather unlikely that more than 1-2 events
among the observed ones can be due to these populations, unless
there are a lot of brown dwarfs, but these would have been
discovered in the same way as in the case of the LMC-5 event.

\subsection{Lenses in the Galactic Spheroid}
Here we examine microlensing events by the stellar halo, i.e. the
so-called {\it Galactic spheroid}. Looking at high latitudes far
above the disk, there is a population of globular clusters, old
metal-poor stars or high-velocity stars, supported by a large
velocity dispersion, $\sigma_{\mathrm{sp}} \simeq (90-120)
\,\mathrm{km \, sec^{-1}}$ (Carrol \& Ostlie 1996; Roulet \&
Mollerach 1997). The spheroid, approximately spherical, provides
an important contribution to the size of the central part of the
Milky Way, while its radial extension is smaller than that of the
halo. Dynamical measurements and infrared surveys predict for the
spheroid a mass roughly equal to $5-7 \times
10^{10}\mathrm{M}_{\sun}$, larger than that estimated on the basis
of star counts. This suggests that the spheroid must be heavy and
mostly dark (Caldwell \& Ostriker 1981; Bahcall et al. 1983).

Giudice et al. (1994) have already discussed in some detail the
possibility to have lenses in the spheroid. Using five different
models based on the works of Caldwell \& Ostriker (1981), Rohlfs
and Kreitschmann (1988), Bahcall et al. (1983), they estimated the
main microlensing quantities: an optical depth in the range
$(3.1-4.9)\,\times 10^{-8}$, an event rate in the range
$(0.01-0.16)/\sqrt{\mu} \,\frac{events}{yr\, 10^{6} stars}$, an
average event duration in the range $(66-71)/\sqrt{\mu}$ days,
where $\mu$, as usual, is the lens mass in solar mass units.

Here we adopt the same simplified density profile, which fits well
the Ostriker and Caldwell model as that in Roulet \& Mollerach
(1997), given by
\begin{eqnarray}\label{spheroid}
\frac{dn}{d\mu}&=&H_{\mathrm{sp}}\frac{dn_{0}}{d\mu}= \\ &=&
\left(\frac
{\sqrt{R_{0}}+\sqrt{R_{\mathrm{c}}}}
{\left(R_{0}^{2}+D_{\mathrm{ol}}^{2}-2R_{0}D_{\mathrm{ol}}\cos{\vartheta}
\right)^{1/4}+\sqrt{R_{\mathrm{c}}}}
\right)^{7}, \nonumber
\end{eqnarray}
where $R_{\mathrm{c}}\simeq0.17 \, \mathrm{kpc}$ is the core
radius (Roulet \& Mollerach 1997) and $\vartheta=82^{\circ}$ is
the angle between the line of sight of the LMC and the direction
of the Galactic center. In this case,
using a local density equal to
$\rho_{0}^{\mathrm{sp}}=1.5\times10^{-3}\,\mathrm{M_{\sun}\,
pc^{-3}}$, our estimate for the optical depth is $\tau=4.29 \times
10^{-8}$, (see Table \ref{G-Tau}) which is in good agreement with the
results of Giudice et al. (1994) and lies well within the range
they found.

Also in this case we used two different IMFs to estimate the
number of events and their mean duration. The results are reported
in Table \ref{G-Nev-T} from which we see that the average duration
is somewhat higher than the observed value, unless the dispersion
velocity is closer to $120 \,\mathrm{km \,sec^{-1}}$ (or if the
IMF goes well into the brown dwarf region). In this latter case
some (2 to 6 events) among the MACHO longer duration events, close
to about 40-50 days, could be due to lenses in the spheroid. This
result is still in a good agreement with Giudice et al. (1994).
Similarly, we expect at most 1 to 2 EROS events to belong to the
spheroid. Clearly, given the calculated mean event duration and
number of expected events we see that it is well plausible that
some of the observed events are due to lenses located in the
spheroid, however, certainly not all of them.
%
\subsection{Lenses in the Galactic Halo}
\label{Lenses in the Galactic Halo}
At last we estimated the microlensing quantities in the hypothesis
that the lenses are MACHOs located in the Galactic halo (Jetzer
1996). The MACHO number density distribution that we used is now
\begin{eqnarray}
\frac{dn}{d\mu}&=&H_{\mathrm{halo}}\frac{dn_{0}}{d\mu}= \\ &=&
\frac{R_{\mathrm{c}}^{2}+R_{0}^{2}}{R_{\mathrm{c}}^{2}+R_{0}^{2}+
D_{\mathrm{ol}}^2-2D_{o\mathrm{l}}
R_{\mathrm{c}}\cos{\vartheta}}\frac{dn_{0}}{d\mu}\nonumber,
\end{eqnarray}
where $\vartheta=82^{\circ}$ is again the angle between the line
of sight and the direction of the Galactic center and
$R_{\mathrm{c}}$ is the core radius of the dark matter, that is
also uncertain. Bahcall et al. (1983) estimated $R_{\mathrm{c}}$
equal to 2 kpc, while Caldwell \& Ostriker (1981) suggested 7.8
kpc. Alcock et al. (2000a) used instead 5 kpc. Here, following
Primack et al. (1988), we use a core radius equal to 5.6 kpc. For
$dn_{0}/d\mu$, we used a delta function, Eq. (\ref{delta}), with
$\rho_{0}^{\mathrm{H}}=7.9\times 10^{-3} \,
\mathrm{M_{\sun}\,pc^{-3}}$.

Fixing the usual halo value for the velocity dispersion,
$\sigma_{\mathrm{H}}=210 \, \mathrm{km\, sec^{-1}}$, we estimated
the optical depth, the number of events and the mean time
duration. The results are tabulated in Tables \ref{G-Tau} and
\ref{G-Nev-T}. For $\mu=0.5\,\mathrm{M_{\sun}}$ (as an
illustration) we expect for a halo made entirely of MACHOs 61
events using MACHO efficiency $\varepsilon=0.4$, and 76 events
using MACHO efficiency $\varepsilon=0.5$, respectively. Similarly
for EROS2, where we use an efficiency equal to $\varepsilon=0.12$,
we expect 19 events. The expected mean event duration is $50
\,\mathrm{days}$ for $\mu=0.5$.

For the optical depth we get the well-known value of $\approx
5\times 10^{-7}$, while the estimated optical depth for a halo
white dwarf population is $\tau=0.13\times 10^{-8}$ (Koopmans \&
Blandford, 2001).
%

\section{Mean Mass}
%
From the previous discussion it is clear that almost certainly not
all the observed events (in particular among the ones detected by
the MACHO collaboration) originate from the same lens population.
Accordingly, it is difficult to estimate a mean value for the
mass, since we do not know in general to which population each
event belongs. In the following we will consider different
possibilities of how to divide the events among the various
populations and so get estimates of the masses using the mass
moment method. Clearly this has to be considered as an
illustration, nonetheless it allows to get some insight and also
to get some upper and lower limits which can already be useful.

The five EROS events, due to their positions (see Fig.
\ref{Mfields}) and durations, can hardly contain self-lensing and
Galactic disk lenses. So most of them if not all (with perhaps the
possibility of having 1 or 2 events due to the spheroid) are
MACHOs in the Galactic halo, in which case we find the following
mean mass: $<M>=0.26\,\mathrm{M_{\sun}}$, assuming a spherical
halo model. If the 1 or 2 events with the longer duration are due
to lenses in the galactic spheroid then of course we would get a
smaller average mass for the remaining 3 to 4 halo events.

The MACHO events contain most probably some lenses in the LMC and
at least one event due to a Galactic component. We exclude as
before from our analysis the MACHO events LMC-9 and LMC-22,
whereas LMC-14, is possibly a self-lensing event. We will
consider, as an illustration, four different models by dividing up
the observed events among the various populations, we shall label
them with: M-1, M-2, M-3 and M-4 (see below).

We already saw in section \ref{Self-lensing-mean-mass} that we
expect at least $3$ self-lensing events. For these three events,
fixing the LMC dispersion velocity to $30\,\mathrm{km/sec}$ and
clearly assuming that one of them is the LMC-14 event, we
estimated a minimum (models M-1 and M-3 in which LMC-1, LMC-14 and
LMC-15 are the self-lensing events) and a maximum (models M-2 and
M-4 in which LMC-7, LMC-13 and LMC-14 are the self-lensing events)
mean mass, see Table \ref{mean mass table} and Fig.
\ref{Mean-mass}.
\begin{table}
\centering \caption{Theoretical estimate of the lens mean mass for
microlensing towards LMC assuming the lens to be in different
components.}
\begin{tabular}{cccc}
\hline \hline
$Lensing \,component$&$Model$&$N_{\mathrm{ev}}$&$<M>$ \\
\hline $Self-lensing$&$\begin{tabular}{c}
M-1 \\
M-2 \\
M-3 \\
M-4 \\
\end{tabular}$&$\begin{tabular}{c}
3 \\
3 \\
3 \\
3 \\
\end{tabular}$&$\begin{tabular}{c}
0.11 $\mathrm{M_{\sun}}$ \\
0.43 $\mathrm{M_{\sun}}$ \\
0.11 $\mathrm{M_{\sun}}$ \\
0.43 $\mathrm{M_{\sun}}$ \\
\end{tabular}$ \\
\hline $Halo$&$\begin{tabular}{c}
M-1 \\
M-2 \\
M-3 \\
M-4 \\
\end{tabular}$&$\begin{tabular}{c}
11 \\
11 \\
5 \\
7 \\
\end{tabular}$&$\begin{tabular}{c}
0.37 $\mathrm{M_{\sun}}$ \\
0.25 $\mathrm{M_{\sun}}$ \\
0.23 $\mathrm{M_{\sun}}$ \\
0.17 $\mathrm{M_{\sun}}$ \\
\end{tabular}$ \\
\hline $Spheroid$&$\begin{tabular}{c}
M-1 \\
M-2 \\
M-3 \\
M-4 \\
\end{tabular}$&$\begin{tabular}{c}
0 \\
0 \\
6 \\
4 \\
\end{tabular}$&$\begin{tabular}{c}
- \\
- \\
0.30 $\mathrm{M_{\sun}}$ \\
0.27 $\mathrm{M_{\sun}}$ \\
\end{tabular}$ \\
\hline $Thick \, disk$&$\begin{tabular}{c}
T-1 \\
T-2 \\
\end{tabular}$&$\begin{tabular}{c}
1 \\
1 \\
\end{tabular}$&$\begin{tabular}{c}
0.08 $\mathrm{M_{\sun}}$ \\
0.24 $\mathrm{M_{\sun}}$ \\
\end{tabular}$ \\
\hline
\end{tabular}
\label{mean mass table}
\end{table}

If only the LMC-5 event is a disk event (we assume it to be in the
thick disk), we found that its mean mass varies from $0.08 \,
\mathrm{M_{\sun}}$ for $\sigma_{\mathrm{TD}}=35\, \mathrm{km/sec}$
(model T-1) to $0.24 \, \mathrm{M_{\sun}}$ for
$\sigma_{\mathrm{TD}}=60\, \mathrm{km/sec}$, (model T-2), see
Table \ref{mean mass table} (assuming instead a velocity of
$20~\mathrm{km/sec}$ as mentioned by the MACHO collaboration
(Alcock et al. 2001c) we find with the mass moment method a mass
of 0.03 $\mathrm{M}_{\sun}$, which compares quite well with the
value inferred by the MACHO team).

For the remaining MACHO events we considered two situations.
\begin{enumerate}
\item They are all due to lenses in the Galactic halo.
The corresponding mean mass is: $<M>=0.37 \,
\mathrm{M_{\sun}}$ using the events LMC-4, LMC-6, LMC-7, LMC-8, LMC-13,
LMC-18, LMC-20, LMC-21, LMC-23, LMC-25, LMC-27 (model M-1)
(mean duration: 39.4 days); $<M>=0.25 \,
\mathrm{M_{\sun}}$ using the events LMC-1, LMC-4, LMC-6, LMC-8, LMC-15,
LMC-18, LMC-20, LMC-21, LMC-23, LMC-25, LMC-27 (model M-2)(mean
duration: 33.4 days).
\item Otherwise, they are due in part to lenses in the spheroid and
in part to lenses in the halo. We divided the events between these
two component preferring for the spheroid the events with longer
durations. We considered again two situations by assuming that in
one case 6 events are due to the spheroid and in the second case
only 4. In the model M-3 we estimated the mean mass for the halo
by using the events LMC-4. LMC-8, LMC-18, LMC-20 and LMC-27 (mean
duration: 30.9 days), while for the spheroid we used the events
LMC-6, LMC-7, LMC-13, LMC-21, LMC-23 and LMC-25 (mean duration:
46.5 days). In the model M-4 the mean mass for the halo was
obtained using the events LMC-1, LMC-4, LMC-8, LMC-15, LMC-18,
LMC-20 and LMC-27 (mean duration: 27.2 days), while for the
spheroid we used LMC-6, LMC-21, LMC-23 and LMC-25 (mean duration:
40.1 days). All the results are shown in Table \ref{mean mass
table}.
\end{enumerate}
We see that for the MACHOs in the Galactic halo we find a mean
mass in the range $(0.17-0.37) \, \mathrm{M_{\sun}}$, which is in
agreement with the previous value of $0.26 \, \mathrm{M_{\sun}} $
obtained from the EROS data. It has to be noticed that these
values are for a standard spherical halo, thus if we consider halo
models with a flattening or with anisotropies in the velocity
space the above values can also vary substantially and in
particular decrease (De Paolis et al. 1996; Grenacher et al.
1999).

For the LMC halo we expect a lens mean mass in the range
$(0.02-0.16)\mathrm{M_{\sun}}$.
%

\section{MACHO mass fraction in the Galactic halo}
%
Another important quantity to be determined is the fraction $f$ of
the local dark mass density detected in form of MACHOs in the
Galactic halo, which is given by (De R\'{u}jula et al. 1991):
\begin{equation}
f \equiv \frac{M_{\sun}}{\rho_{0}} <\mu^{1}>\, \simeq 127 \,
\mathrm{pc^{3}}<\mu^{1}>.
\end{equation}
Using the values given by the MACHO collaboration for their 5.7 years
data, we estimated $f$ for each of the four models considered
before and also for the EROS2 data assuming that all their events
are due to MACHOs in the halo. The results, obtained by assuming
again a standard spherical halo model as in Sec. \ref{Lenses in
the Galactic Halo}, are reported in Table \ref{fraction}.
\begin{table}
\centering \caption{Fraction of the local dark mass density in
form of MACHOs in the Galactic halo, which we get using MACHO and
EROS data.}
\begin{tabular}{cccc}
\hline \hline
$$&$N_{\mathrm{ev}}$&$\varepsilon$&$f$ \\
\hline $\mbox{M-1}$&$11$&$\begin{tabular}{c}
0.4 \\
0.5 \\
\end{tabular}$&$\begin{tabular}{c}
13\% \\
11\% \\
\end{tabular}$ \\
\hline $\mbox{M-2}$&$11$&$\begin{tabular}{c}
0.4 \\
0.5 \\
\end{tabular}$&$\begin{tabular}{c}
11\% \\
9\% \\
\end{tabular}$ \\
\hline $\mbox{M-3}$&$5$&$\begin{tabular}{c}
0.4 \\
0.5 \\
\end{tabular}$&$\begin{tabular}{c}
5\% \\
4\% \\
\end{tabular}$ \\
\hline $\mbox{M-4}$&$7$&$\begin{tabular}{c}
0.4 \\
0.5 \\
\end{tabular}$&$\begin{tabular}{c}
6\% \\
5\% \\
\end{tabular}$ \\
\hline
$\mbox{EROS} \, 2$&$4$&$0.12$&$12\%$ \\
\hline
\end{tabular}
\label{fraction}
\end{table}

We see that for the models we considered the fraction of halo dark
matter in form of MACHOs varies between 5\% and 13\%. Moreover, when
neglecting contributions from the spheroid (models M-1 and M-2)
the values we get from MACHO and EROS collaboration are
practically similar. Clearly, if some events are indeed due to the
spheroid (as in models M-3 and M-4) we should then take this
component also into account for the EROS data. In which case the
fraction we find from the EROS experiment gets accordingly smaller
and is then again in good agreement with the MACHO value.
%

\section{Asymmetry of the observed events in the LMC}
A striking feature which comes out by examining the positions of
the events found by the MACHO and EROS collaborations is that
there is a clear asymmetry in the spatial distribution of the
events with respect to the bar major axis. The events are
concentrated along the extension of the bar and in the south-west
side of LMC (Fig. \ref{Mfields}). In particular, from MACHO data
we have 12 events located in the south-west side and 5 events
located in the north-east side. The ratio is just $12/5=2.4$. To
study the nature of this asymmetry, we divided the 27 MACHO fields
that we selected in two groups. In the first group there are the
fields located above the bar major axis and in the second those
below. Four fields (11, 12, 47, 79) due to their location were
split among the two groups. We added the products obtained by
multiplying the number of stars, $N_{\star}$, by the observing
time, $T_{\mathrm{obs}}$, that is clearly proportional to the
number of observations pro field reported by the MACHO
collaboration (Alcock et al. 2000a), weighted by the corresponding
event rate for self-lensing $\Gamma$ for each of the fields in one
group. The ratio of the two sums turns to be close to 1, more
precisely
\begin{equation}
\frac
{\left(\sum_{i=1}^{16}N_{\star,i}\,T_{\mathrm{obs},i}\,
\Gamma_{i}\right)_{sw}}
{\left(\sum_{i=1}^{11}N_{\star,i}\,T_{\mathrm{obs},i}\,
\Gamma_{i}\right)_{ne}}=0.96.
\end{equation}
This result is obtained using the model $1a$ and IMF-2, although
we would get a similar value also by using the other models. If
the events are due to self-lensing we would expect the ratio of
the events to be close to 1 rather than being 2.4. Clearly, the
same holds if the events are due to MACHOs in the halo or in one
of the considered galactic populations. A possible explanation
might be that the lenses are located in the halo of the LMC which
would not be spherical and perhaps more elongated towards our
galaxy, or an alternative is that the MACHOs in our halo are not
uniformly distributed.
%

\section{Discussion}
%
As mentioned at the beginning the issue of the location and nature
of the objects which act as lenses in the observed microlensing
events is still an open problem, which can possibly be solved once
more events will be available. The number of events found by the
experimental collaborations are too few in comparison with that
predicted for a halo composed entirely by MACHOs. In the last
years several possible explanations of the experimental data have
been proposed, but they are all not definitive and not completely
satisfactory. One possibility might be that there are more MACHOs
in the halo, but that, for instance, they are associated with gas
cloud (Bozza et al. 2002), which would then produce non-achromatic
events, that due to the present selection criteria have not been
considered. On the other hand, one should also consider the
possibility that some events might not be due to microlensing at
all. This obviously underlines the fact that the present results
have to be taken with care and that more observations are needed
in order resolve this issue.

From the presently few available events and our above discussion
it emerges clearly that, especially for the MACHO events, the
lenses are due to different populations. Some are certainly due to
LMC self-lensing, but this can hardly be the case for all the
observed events especially for the few EROS events.

For the preferred LMC model with a dispersion velocity of about
$30~ \mathrm{km\, sec^{-1}}$ we expect some 3 events among the
MACHO ones (not including the possible self-lensing binary event
LMC-9) to be due to self-lensing. Moreover, with the mass moment
method we find an average mass in the range $(0.1 - 0.5)\,
\mathrm{M_{\sun}}$ for the self-lensing events, which is
consistent with the expectation that the lenses are low mass
stars. The contribution of a LMC halo is, even if it exists, also
very minor of at most 1 event unless we are in the rather strange
situation where the LMC halo is, contrary to our own, made almost
entirely of MACHOs. From the galactic component, thin and thick
disk, we expect roughly $1-3$ events on the MACHO data, but no
contribution on the EROS data. This result seems to be in
agreement with the fact that LMC-5 is a disk event. The inferred
mass is small, though compatible with a low mass star, but clearly
with only one event at disposal one has to take this value just as
indicative.

As a result of our analysis we find that a plausible solution is
that among the MACHO data some $3-4$ events are due to
self-lensing, $1-2$ to the thick disk and for the remaining ones
about half are due to the spheroid and the others to the halo. If
so the optical depth gets contributions from each of these
components, namely: about $2.3 \times 10^{-8}$ from self-lensing,
some $(2-4) \times 10^{-8}$ from the halo, $4 \times 10^{-8}$ from
the spheroid and $4 \times 10^{-8}$ from thin and thick disk. This
way we get a total optical depth of about $(1.2 - 1.5) \times
10^{-7}$ which is in good agreement with inferred value by the
MACHO team of $\tau=1.2_{-0.3}^{+0.4}\times 10^{-7}$. Since the
EROS events are less and given their spatial position, we do not
expect that they get much contribution from self-lensing and the
thick disk so that it is clearly not possible to draw more
conclusions from them. On the other hand assuming that the EROS
events are due to lenses in the halo, with or without some
contribution from the spheroid, leads to a halo mass fraction
which is in reasonable agreement with the corresponding MACHO
value once the self-lensing and the disk events are subtracted.
This way the MACHO and EROS results nicely fit together. Moreover,
the above mentioned asymmetry, if not just due to a statistical
fluctuation, is also not compatible with only self-lensing events.
Given our results it is also clear that one has to take with care
values on the lens mass based on the assumption that all lenses
belong to just one population. Clearly, once more data will be
available, by using also the methods outlined in this paper, it
will be possible to draw more firm conclusions.
%

\begin{acknowledgements}
The authors thank D. Bennett, V. Bozza, N. Dalal and A. Milsztajn
for their communications and useful suggestions. This work is
partially supported by the Swiss National Science Foundation.
\end{acknowledgements}
%
\appendix
%
\section{Coordinate transformations}
\label{Aa}
To quantify $\rho_{\mathrm{d}}$ or $\rho_{\mathrm{b}}$, we
introduce the coordinate system $\lbrace x_{0},y_{0},z_{0}\rbrace$
which has the origin in the center of the LMC at $\lbrace l,b,D
\rbrace=\lbrace l_{0},b_{0},D_{0}\rbrace$, and has $z_{0}$-axis
toward the observer, $x_{0}$-axis anti parallel to the galactic
longitude axis, and $y_{0}$-axis parallel to the galactic latitude
axis. $D_{0}\simeq50 \, \mathrm{kpc}$ is the distance between the
center of the LMC and the observer, while $D$ is the generic
observer-lens or observer-source distance. $\lbrace l_{0},
b_{0}\rbrace$ are the galactic coordinates of the center of the
LMC. The coordinate transformations are given by
\begin{eqnarray}\label{first trasformations}
&x_{0}&=-D\cos{b}\sin{(l-l_{0})}\nonumber
\\ &y_{0}&=D\sin{b}\cos{b_{0}}-D\cos{b}\sin{b_{0}}\cos{(l-l_{0})}
\\ &z_{0}&=D_{0}-D\cos{b}\cos{b_{0}}\cos{(l-l_{0})}-D\sin{b}\sin{b_{0}}.
\nonumber
\end{eqnarray}
The coordinate system of the LMC disk $\lbrace x,y,z\rbrace$ is
the same orthogonal system as $\lbrace x_{0},y_{0},z_{0}\rbrace$,
except that it is rotated around $z_{0}$-axis by the position
angle $\phi$ counterclockwise and around the new  $x$-axis by the
inclination angle $i$ clockwise. The coordinate transformations
are given by
\begin{eqnarray}\label{second trasformations}
&x&=x_{0}\cos{\phi}+y_{0}\sin{\phi}\nonumber
\\ &y&=-x_{0}\sin{\phi}\cos{i}+y_{0}\cos{\phi}\cos{i}-
z_{0}\sin{i}
\\ &z&=-x_{0}\sin{\phi}\sin{i}+y_{0}\cos{\phi}\sin{i}+
z_{0}\cos{i} .\nonumber
\end{eqnarray}
%


\bibliographystyle{aa}

\end{document}